
%
%
%
\input harvmac
\overfullrule=0mm
\def\gl{\lambda}
\def\gL{\Lambda}
\def\d{ d}
\def\tr{{\rm tr}}
\def\c{{\rm ch}}

\def\encadre#1{\vbox{\hrule\hbox{\vrule\kern8pt\vbox{\kern8pt#1\kern8pt}
\kern8pt\vrule}\hrule}}
\def\encadremath#1{\vbox{\hrule\hbox{\vrule\kern8pt\vbox{\kern8pt
\hbox{$\displaystyle #1$}\kern8pt}
\kern8pt\vrule}\hrule}}

%
\def\frac#1#2{\scriptstyle{#1 \over #2}}		
\def\inv#1{\scriptstyle{1 \over #1}}

\def\d{\partial}

\def\dd#1#2{{\partial #1 \over \partial #2}}

%
%

%
\def\({ \left( }
\def\){ \right) }
%


\def\IR{\relax{\rm I\kern-.18em R}}
\font\cmss=cmss10 \font\cmsss=cmss10 at 7pt
\def\IZ{\relax\ifmmode\mathchoice
{\hbox{\cmss Z\kern-.4em Z}}{\hbox{\cmss Z\kern-.4em Z}}
{\lower.9pt\hbox{\cmsss Z\kern-.4em Z}}
{\lower1.2pt\hbox{\cmsss Z\kern-.4em Z}}\else{\cmss Z\kern-.4em Z}\fi}
\def\inbar{\,\vrule height1.5ex width.4pt depth0pt}
\def\IB{\relax{\rm I\kern-.18em B}}
\def\IC{\relax\hbox{$\inbar\kern-.3em{\rm C}$}}
\def\ID{\relax{\rm I\kern-.18em D}}
\def\IE{\relax{\rm I\kern-.18em E}}
\def\IF{\relax{\rm I\kern-.18em F}}
\def\IG{\relax\hbox{$\inbar\kern-.3em{\rm G}$}}
\def\IH{\relax{\rm I\kern-.18em H}}
\def\II{\relax{\rm I\kern-.18em I}}
\def\IK{\relax{\rm I\kern-.18em K}}
\def\IL{\relax{\rm I\kern-.18em L}}
\def\IM{\relax{\rm I\kern-.18em M}}
\def\IN{\relax{\rm I\kern-.18em N}}
\def\IO{\relax\hbox{$\inbar\kern-.3em{\rm O}$}}
\def\IP{\relax{\rm I\kern-.18em P}}
\def\IQ{\relax\hbox{$\inbar\kern-.3em{\rm Q}$}}
\def\IGa{\relax\hbox{${\rm I}\kern-.18em\Gamma$}}
\def\IPi{\relax\hbox{${\rm I}\kern-.18em\Pi$}}
\def\ITh{\relax\hbox{$\inbar\kern-.3em\Theta$}}
\def\IOm{\relax\hbox{$\inbar\kern-3.00pt\Omega$}}


\def\d{{\rm d}}

\def\oh{{1\over 2}}


\def\GD{\Delta}
\def\Gth{\theta}
\def\Gl{\lambda}\def\GL{\Lambda}

\def\GS{\Sigma}


\def\d{{\rm d}}

\def\Gthd{\Gth_{\textstyle .} }
\def\fd{f_{\textstyle .} }

\def\KK{Kontsevich}

\Title{SPhT 92/073}
{{\vbox {
\bigskip
\centerline{ Polynomial Averages in the Kontsevich Model}
}}}

\bigskip

\centerline{P. Di Francesco,}
\bigskip\centerline{C. Itzykson}

\bigskip
\centerline{and}
\bigskip
\centerline{J.-B. Zuber}
\bigskip
\centerline{ \it Service de Physique Th\'eorique de Saclay
\footnote*{Laboratoire de la Direction des Sciences et
de la Mati\`ere du Commissariat \`a l'Energie Atomique.},}
\centerline{ \it F-91191 Gif sur Yvette Cedex, France}

\vskip .2in

\noindent
We obtain in closed form averages of polynomials, taken
over hermitian matrices with the Gaussian measure
involved in the Kontsevich integral, and prove a conjecture
of Witten enabling one to express analogous averages with
the full (cubic potential) measure, as derivatives of the
partition function with respect
to traces of inverse odd powers of the external argument.
The proofs are based on elementary algebraic
identities involving a new set  of
invariant polynomials of the linear group,
closely related to the general
Schur functions.

\Date{06/92}
%

\lref\KON{M. \KK,
\it Intersection theory on the moduli space of curves\rm,
Funk. Anal.\& Prilozh., {\bf 25} (1991) 50-57\semi
\it Intersection theory on the moduli space of curves and the matrix Airy
function\rm,
lecture at the Arbeitstagung, Bonn, June 1991 and
Bonn preprint MPI/91-77.}
\lref\Wun{E. Witten,
\it Two dimensional gravity and intersection theory on moduli space\rm,
Surv. in Diff. Geom. {\bf 1} (1991) 243-310.}
\lref\WIT{E. Witten,
\it On the \KK\ model and other models of two dimensional gravity\rm,
preprint IASSNS-HEP-91/24}
\lref\Wdep{E. Witten,
\it The $N$ matrix model and gauged WZW models\rm,
preprint IASSNS-HEP-91/26, to appear in Nucl. Phys. B. }
\lref\Wtr{E. Witten,
\it Algebraic geometry associated with matrix models of two dimensional
gravity, \rm
 preprint IASSNS-HEP-91/74.}
\lref\CJB{ C. Itzykson and J.-B. Zuber,
\it Combinatorics of the Modular Group II: The \KK\ integrals. \rm,
to appear in Int. J. Mod. Phys. }
\lref\INT{Harish-Chandra,
{\it Differential operators on a semisimple Lie algebra},
Amer.J.Math. {\bf 79} (1957) 87-120 \semi
C. Itzykson and J.-B. Zuber,
\it The planar approximation II,\rm\  J. Math. Phys. {\bf 21} (1980) 411-421.}

\newsec{Introduction}

\noindent In their papers on intersection theory on moduli spaces of Riemann
surfaces, Witten \WIT\ and \KK\ \KON\
discussed  certain  identities on matrix integrals.
We provide here algebraic proofs for these statements which read as
follows.

Let $X$, $Y$, $\gL$ ... denote $N\times N$ hermitian matrices.
For $\gL$ positive definite, hence ${\gL}^{-1}$ well defined,
introduce the measure
\eqn\measure{\eqalign{
 \d \mu_{{\gL}}^{(N)}(Y)&=(2 \pi)^{-N^2/2}
\prod_{i=1}^N \d Y_{ii} \prod_{1 \leq i<j \leq N}
\d { Re} Y_{ij} \d { Im} Y_{ij} \exp{-{1 \over 2} \tr \gL Y^2 } \cr
&= {\rm d}Y  \exp{-{1 \over 2} \tr \gL Y^2 } \cr}}
\medskip

\noindent{ \bf Proposition (K)}  For any polynomial $P$ in the traces
of odd powers of $Y$ (``odd traces'' for short),
there exists a polynomial $Q$ in odd traces of
${\gL}^{-1}$,such that, independently of $N$ large enough
\eqn\K{ \langle P \rangle_{(N)} (\gL^{-1}) =
{{\int \d \mu_{\gL}^{(N)}(Y) P(Y) } \over { \int \d \mu_{\gL}^{(N)}(Y) }}
=Q({\gL}^{-1}) }
\medskip

In section 2 we exhibit more precisely the surjective map $K$ : $P \to Q$
(proposition (K'))
which turns out to be defined over $\IQ$, the rationals, and obtain
its kernel.
Upon applying Wick's theorem to the computation of $\langle P \rangle$,
i.e. performing a ``fat graph" expansion \WIT \KON, the proposition
amounts to intricate algebraic identities, since each graph contributes
a symmetric rational function of the eigenvalues of $\gL$.
In a previous paper \CJB,
two of the authors proved what amounts to a
special case of this conjecture ($P$ was a polynomial in $\tr Y^3$
only) by a detailed but painful analysis, which admitted however
a generalization to more general integrals not considered here.

It follows from (K) that the integral
\eqn\kont{ \Xi_{N}(\gL^{-1})= {{ \int \d \mu_{\gL}^{(N)}(Y)
\exp{{i \over 6} \tr Y^3} } \over { \int \d \mu_{\gL}^{(N)}(Y) }}
= \langle \exp{{i \over 6} \tr Y^3 } \rangle_{(N)}(\gL^{-1}) }
admits as $N \to \infty$ an asymptotic expansion $\Xi(\theta_{
\scriptstyle{.}})$,
each term of
which is a polynomial in the normalized odd traces ($(-1)!!=1$)
\eqn\nort{ \theta_{2k+1}= -{2 \over 2k+1} \tr {\gL}^{-2k-1} }
which become independent variables.
For a more accurate definition see \CJB. Similarly for $P$ a polynomial
in the odd traces of $Y$, the quantity
\eqn\avkon{ \ll P \gg_{(N)} = \langle P(Y) \exp{ {i \over 6} \tr Y^3}
\rangle_{(N)}(\gL^{-1}) }
admits an asymptotic expansion $\ll P \gg (\theta_{\scriptstyle{.}})$,
each term of which is independent of $N$ for $N$ large enough.
We then have the second proposition conjectured by Witten and
discussed in section 3
\medskip

\noindent{ \bf Proposition (W)} For each $P$ as above there exists
a polynomial $R$ in the derivatives $\partial_{\theta_{\scriptstyle{.}}} \equiv
\{ \partial_{\theta_1},\partial_{\theta_3},... \}$ such that
\eqn\W{ \ll P \gg (\theta_{\scriptstyle{.}})=
R(\partial_{\theta_{\scriptstyle{.}}} )
\  \Xi(\theta_{\scriptstyle{.}}) }
\medskip

The invertible mapping $P \leftrightarrow R$, defined over $\IQ$,
is given explicitely by proposition (W').

The idea  underlying these proofs
is simple enough as it reduces to compare
calculations for matrices of size differing by a finite amount.
To make this rigorous, we have to follow an indirect path which
unfortunately  tends to obscure the proofs in a maze of cumbersome
notations. On our way we are led to introduce a set of polynomials
denoted $f_{\scriptstyle{.}}(\theta_{\scriptstyle{.}})$, in infinitely
many variables $\theta_{\scriptstyle{.}}\equiv
\{ \theta_1,\theta_3,..,\theta_{2k+1},...\}$, closely related to the
generalized Schur functions, in terms of which our results are most
simply expressed (see eqns. (2.18) and (2.27)).
We tabulate the first few $f_{\scriptstyle{.}}$'s as well as some
expressions obtained in the text (tables I-IV).

The reader might rightly wonder about the meaning of such results.
The answer is that both (K) and (W) enable one to
construct general ``observables", or topological invariants,
in the combinatorial treatment of moduli spaces \WIT\ \KON.

This work was prompted by questions raised by E. Witten
and M. Kontsevich
\ref\WKP{ E. Witten and M. Kontsevich, private
communications.}.

\newsec{ Proof of proposition (K)}

\subsec{ Preparation}

\noindent
For $X$ a generic $N \times N$ hermitian matrix with eigenvalues
$x_1,x_2,...,x_N$, and $u$ a complex variable, the rational function
\eqn\determ{ F_1(u;X)=\det {{1 - uX} \over {1+uX}} }
admits for $u$ in the disk $D(X) \equiv \{ |u|<{1 \over { \sup_i |x_i|}}
\}$ a convergent power series
\eqn\powser{ F_1(u;X)= \sum_{k=0}^{\infty} u^k p_k(X) }
Each $p_k$ is a polynomial in the odd traces of $X$, normalized in this
section as
\eqn\northeta{ \theta_{2n+1}(X) = -2 { \tr X^{2n+1} \over {2n+1}},}
homogeneous of degree $k$, by assigning degree $2n+1$ to
$\theta_{2n+1}$, as follows from identifying
\eqn\sersch{ F_1(u;X)=\exp \big(-2\sum_{n=0}^{\infty} u^{2n+1}
\tr{ X^{2n+1} \over {2n+1}} \big)  = \sum_{k=0}^{\infty} u^k p_k(X)}
Considered as a polynomial in the infinitely many variables $\Gthd$
we have therefore
\eqn\schur{ p_k(\Gthd)=\sum_{\nu_j \geq 0,j \ {\rm odd} \atop
k=\nu_1+3 \nu_3+...} \prod_{j=1,3,...}
{{\theta_j^{\nu_j} }\over
{\nu_j !} } }
By abuse of language we will call these functions Schur polynomials,
although they are obtained from the standard ones by setting
all the even variables $\theta_{2j}$ to zero.
We also extend the standard definition by setting $p_k=0$ for $k<0$.

More generally, for $k_1, k_2,...,k_n$ non negative integers
\eqn\defchi{ \c_{k_1,...,k_n}(\Gthd) = \left\vert
\matrix{ p_{k_1-n+1} &
p_{k_1-n+2} &...&p_{k_1} \cr
p_{k_2-n+1} & p_{k_2-n+2}&...&p_{k_2} \cr
.&.&...&. \cr
.&.&...&. \cr
p_{k_n-n+1}&p_{k_n-n+2}&...&p_{k_n} \cr} \right\vert (\Gthd) }
is a polynomial in the odd traces $\Gthd$, antisymmetric in its
indices, of degree
\eqn\degree{ \d_{\{ k \}} = \sum_{r=1}^n k_r -{{n(n-1) } \over 2} }
which has to be non negative for $\c_{\textstyle{.}}$ to be non vanishing
(observe that $\c_{k_1,...,k_n,0}=\c_{k_1-1,...,k_n-1}$).
Again we extend the definition by requiring $\c_{\textstyle{.}}$ to vanish
if any of its indices is negative.
Expressed in terms of $\Gthd$'s, $\c_{\textstyle{.}}$ is akin to an ordinary
character of the linear group when we let the even $\theta$'s vanish
\foot{ Of course for the standard characters one sets $\theta_n=
\tr {X^n \over n}$ for any positive $n$.}.

We introduce the following generating function
\eqn\genfunc{
\eqalign{F_n(u_1,...,u_n;X)&=\phi_n(u_1,...,u_n)
\prod_{1 \leq i<j \leq n} \det {{1-u_i X} \over {1+u_i X}} \cr
\phi_n(u_1,...,u_n)&= \prod_{1 \leq i<j \leq n}
{{u_i- u_j} \over {u_i + u_j}} \ \ \ \ ;\ \phi_1=\phi_0=1 \cr}}
such that
\eqn\gener{ \prod_{1 \leq i<j \leq n}(u_i+u_j)
\, F_n(u_1,...,u_n;X)={1 \over n!}
\sum_{k_1,...,k_n} |u^{k_1},...,u^{k_n}| \ \c_{k_1,...,k_n}(X) }
where the short hand notation stands for
\eqn\shorthand{ |u^{k_1},...,u^{k_n}| = \left\vert \matrix{
u_1^{k_1}&....&u_1^{k_n} \cr
.&....&. \cr
.&....&. \cr
u_n^{k_1}&....&u_n^{k_n} \cr } \right\vert }
and $F_0=1$, $F_n=0$ for $n<0$.
We can then think of $F_n$ as a function antisymmetric in the
$n$ variables $u_i$ and of infinitely many variables $\Gthd$,
by substituting for $\c_{\textstyle{.}}$ in the series expansion its
expression as a polynomial in the odd $\Gthd$'s.
The following enables us to replace in the statement of proposition
(K) any polynomial $P$ in the $\Gthd$'s by a polynomial in
the $\c_{\textstyle{.}}(\Gthd)$.

\noindent{ \bf Lemma 1} Any homogeneous polynomial in the
$\Gthd$'s  admits an expansion in
terms of $\c_{\textstyle{.}}$'s of the same degree with ordered positive
indices.

\def\chit{\chi\dup_{T}}
\def\cht{{\rm ch}\dup_{T}}

Indeed if to $k_1>k_2>...>k_n$ we let correspond the Young tableau $T$
with $k_1-n+1$ boxes in the first row, $k_2-n+2$ in the second ..., the
standard Frobenius duality relation takes in our case the following form
\eqn\frob{ \prod_{j=0}^{\infty} ((2j+1)\theta_{2j+1})^{\nu_{2j+1}}=
\sum_{T;|T|=\sum_{j} (2j+1)\nu_{2j+1}} \chit([1^{\nu_1} 3^{\nu_3}...])
\cht(\Gthd) }
So on the one hand the definition \defchi\
of $\c_{\textstyle{.}}$ yields its expression in
terms of $\theta$'s through Schur polynomials, and on the other hand
the above relation, where $\chit([1^{\nu_1} 3^{\nu_3}...])$
denotes the character of the symmetric group on $|T|$ symbols
evaluated on the corresponding class (involving only odd cycles), yields
reciprocally an expression of any monomial in the $\theta$'s as
combination of $\c_{\textstyle{.}}$'s.
We remark that all coefficients involved are rational since
$\chit$ takes only integer values
\foot{ The lemma does not imply that the $\c_{\textstyle{.}}(\Gthd)$ are
linearly independent for if ${\widetilde T}$ is the Young tableau
dual to $T$, $\c_{{\widetilde T}}(\Gthd)=\cht(\Gthd)$
as a consequence of the  fact that $\chit$
and $\chi_{{\widetilde T}}$,
which differ only by the signature of the permutation, are
equal on classes involving only odd cycles, the latter corresponding to
even permutations. Unfortunately, these are not the only relations
on the $\c_T(\Gthd)$ which form an overcomplete system
of generators. For a better choice, see below.}.

To understand the origin of the property expressed by (K),
consider the Gaussian integral
\eqn\gauss{ Z_{N}(\gL) = \int d \mu_{\gL}^{(N)}(Y)
= {1 \over {(\det \gL)^{1/2}
\prod_{1 \leq i<j \leq N}(\gl_i+\gl_j) }}}
where $\gl_1,...,\gl_N$ are the positive eigenvalues of $\gL$ and we
use the positive square root of $\det \gL$.
After performing the ``angular " average over the argument $Y$
\INT,
this becomes
\eqn\debob{\eqalign{
{{\widetilde Z}}_N(\gL)&=
\prod_{1 \leq i<j \leq N} (\gl_i - \gl_j)\, Z_N(\gL) \cr
&= {1 \over {(\det \gL)^{1/2}}} \prod_{1 \leq i<j \leq N}
{{\gl_i - \gl_j} \over {\gl_i +\gl_j}} \cr
&= {{(-1)^{N(N-1) /2}} \over N!} \int \prod_{i=1}^N
{dy_i \over \sqrt{ 2 \pi}}
\det [ e^{-{{\gl_l y_m^2} \over 2}} ]_{1 \leq l,m \leq N}
\prod_{1 \leq i<j \leq N} {{y_i - y_j} \over {y_i+y_j}} \cr }}
The integral over the eigenvalues $y_i$ of $Y$ is well defined
since for $i \neq j$ as $y_i+y_j \to 0$ the determinant in the
numerator vanishes, whereas the exponential factors (for $\gL>0$)
ensure convergence at infinity. Both sides are antisymmetric
in the $\gl$'s and up to factors we see that the (antisymmetrized)
Gaussian transform of $\prod_{i<j} (y_i -y_j)/ (y_i + y_j)$
is a similar expression in the $\gl^{-1}$'s.

This suggests to compare $Z_{N+n}$ to $Z_N$ with an argument
of the form $\gL' \oplus \gL$ where $\gL'$ is an $n \times n$
matrix with eigenvalues $\gl_1',...,\gl_n'$ and $\gL$ as above.
We can also split the integration variable $Y' \oplus Y$ in the diagonal
form occurring in \debob.
We integrate separately over $Y$ and $Y'$ to get
averages over functions $F_n$ of the argument $Y$ in terms of
similar functions of $\gL^{-1}$.
Both admit expansions in odd traces of $\gL'^{-1}$
and are at the origin of
the property expressed in proposition (K).

For analytic reasons it is however difficult to carry out this program
directly. Therefore we take an indirect route  based on the same
idea which can be summarized as follows.

We have first traded polynomials in odd traces (of $Y$ or $\gL^{-1}$)
for linear combinations of $\c_T(\Gthd)$.
In a second step we will substitute for $\c_T(\Gthd)$
an equivalent complete set denoted $\fd(\Gthd)$
indexed by {\it positive} integers, also antisymmetric
in its indices, and shall prove the main result of this section
\medskip

\noindent{\bf Proposition (K')}
Any polynomial in odd traces admits a unique expansion in
terms of $\fd$'s and vice versa.
Moreover, independently of $N$ large enough
\eqna\main{
$$\eqalignno{
&\encadremath{
\ \
\langle f_{k_1,...,k_n}\rangle_{(N)} (\gL^{-1})=0 \ \  {\rm if} \  {\rm at}\
{\rm least}
\ {\rm one} \ {\rm of } \ {\rm the}
\ k_i  \  {\rm  is}\ {\rm odd}\ \ \ \ \ \ \  } &\main a \cr
&\encadremath{
\langle f_{2k_1,...,2k_n}\rangle_{(N)}(\gL^{-1})=
\prod_{s=1}^n (2k_s-1)!!(-1)^{k_s} \ f_{k_1,...,k_n}(\Gthd(\gL^{-1}))}
&\main b\cr}$$ }
\medskip

We have therefore (i) to define the $f$'s (ii) to show their
equivalence with the $\c$'s and linear independence
(iii) to obtain the integrals \main{a-b}.
Completing these three steps will prove the proposition.

\subsec{ Definition of the $f_{.}(\Gthd)$}

For $X$ hermitian $N \times N$, let $x_1,...,x_N$ be its eigenvalues
assumed all distinct.
We define $X_a$ as hermitian $N-1 \times N-1$, with eigenvalue
$x_a$ omitted, similarly for $X_{a_1,a_2,...}$.

As a meromorphic function in the variable $u_1^{-1}$,
$F_n(u_1,...,u_n;X)$ defined in \genfunc\
can be expanded as a sum over its simple poles
plus a contribution at infinity (arguments with  a hat
are to be omitted)
\eqn\decelsim{
\eqalign{
F_n(u_1,...,u_n;X)&= (-1)^{n-1} F_{n-1}({\hat u_1},u_2,...,u_n;X) \cr
&-2\sum_{a=1}^N {{u_1 x_a }\over { 1+u_1 x_a}}
{{ F_{n-1}({\hat u_1},u_2,...,u_n;X_a)} \over {F_1(x_a^{-1};X_a)}} \cr
&+2\sum_{l=2}^n (-1)^l {{u_1} \over {u_1+u_l}}
F_{n-2}({\hat u_1},u_2,...,{\hat u_l},...,u_n;X) \cr}}
Upon iteration, this yields
\eqn\iterf{
\eqalign{
F_n(u_1,...,u_n;X) &= \sum_{r=0}^{\infty} (-2)^r
\sum_{1 \leq i_1<..<i_r \leq n}
\phi_{n-r}(u_1,..,{\hat u_{i_1}},..,{\hat u_{i_r}},...,u_n)
(-1)^{{\cal P}_{\{i\}}} \cr
&\times \sum_{1\leq a_1<...,a_r \leq N}
{{(-1)^{r(r-1)/2} } \over
{F_r(x_{a_1}^{-1},...,x_{a_r}^{-1};X_{a_1,...,a_r})}}
\det\left[ {{u_{ i_s} x_{a_t} } \over
{1+ u_{i_s} x_{a_t}}} \right]_{1 \leq s,t \leq r} \cr}}
Here $(-1)^{{\cal P}_{\{i\}}}$ is the signature of the permutation
$(1,...,n) \to (i_1,..,i_r,1,.., {\hat i_1},..,{\hat i_r},..,n)$, and
$\phi$ is defined in \genfunc.
The first term ($r=0$) on the right hand side is
$\phi(u_1,...,u_n)$ and
\eqn\moref{ {{(-1)^{r(r-1)/2} } \over
{ F_r(x_{a_1}^{-1},...,x_{a_r}^{-1};X_{a_1,...,a_r})}}=
\prod_{1 \leq s<t \leq r} {{x_{a_s}+x_{a_t}} \over {x_{a_s}-x_{a_t}}}
\prod_{1 \leq s \leq r \atop l \in \{1,2,..,{\hat a_1},..,{\hat a_r},..N\}}
{{x_{a_s}+x_l} \over {x_{a_s}-x_l}} }
For $n,k_i>0$ define
\eqn\fdef{ f_{k_1,...,k_n}(X)= 2^n (-1)^{k_1+...+k_n}
\sum_{1 \leq a_1<...<a_n \leq N} {{ (-1)^{n(n-1)/2}}
\over {F_n(x_{a_1}^{-1},...,x_{a_n}^{-1};X_{a_1,...,a_n})}}
\det [x_{a_s}^{k_t}]_{1 \leq s,t \leq n} }
and extend this definition to $n=0$ by setting $f(X)=1$.
These functions (antisymmetric in $k_1,...,k_n$) appear
at first as rational symmetric functions of $X$.
We shall soon see that they are in fact polynomials in
$\Gthd(X)$ and thus still well defined when some
eigenvalues coincide.
Let us insist on the fact that we assume all indices positive,
otherwise let $\fd=0$.
We can now compare the two expansions of
$\prod_{i<j} (u_i+u_j) F_n(u_1,...,u_n;X)$ on the
polydisc $u_i \in D(X)$, namely
\eqn\compchif{
\eqalign{
{1 \over n!} &\sum_{k_1,...,k_n \geq 0} \ |u^{k_1}...u^{k_n}|
\ \c_{k_1,...,k_n}(\Gthd(X))=
\prod_{1 \leq i<j \leq n} (u_i -u_j)+
\sum_{r=1}^n
{1 \over r!} \times \cr
&\times \sum_{k_1,...,k_r >0} f_{k_1,...,k_r}(X)
\sum_{1 \leq i_1<...<i_r \leq n} (-1)^{{\cal P}_{\{i\}}}
\prod_{1 \leq i<j \leq n} (u_i + \epsilon_{ij} u_j)
\left\vert \matrix{ u_{i_1}^{k_1} & ...&u_{i_1}^{k_r} \cr
.&...&. \cr
u_{i_r}^{k_1}&...&u_{i_r}^{k_r} \cr } \right\vert  \cr}}
where
\eqn\psilon{ \epsilon_{ij}= \left\{ \eqalign{ -1& \ {\rm if} \ i,j
\in \{ 1,..,{\hat i_1},...,{\hat i_r},...,n\} \cr
1& \ {\rm otherwise}\cr } \right. }
We need a few extra notations.
Let the generalized antisymmetric Kronecker symbol
be
\eqn\krogen{ \delta_{k_1,...,k_r;m_1,...,m_r}=
\det [\delta_{k_i,m_j}]_{1 \leq i,j \leq r} }
and define a shift operator by
\eqn\shiftop{ g_m \to (P^l g)_m \equiv g_{m+l} \ ,
\ l \in {\bf Z} }
Finally we write $I \cup J$ for a partition of
$\{ 1,...,n\}$ into two disjoints ordered sets
$I\equiv \{ i_1,...,i_{|I|} \}$ and
$J \equiv \{ j_1,...,j_{|J|}\}$, $|I|+|J|=n$, and
denote by $(-1)^{{\cal P}_{I,J}}$
the signature of the permutation
$(1,...,n) \to (i_1,..,i_{|I|},j_1,...,j_{|J|})$.

Identifying the antisymmetric coefficient of
$|u^{k_1}...u^{k_n}|$ in the previous equality
yields the desired linear relation between $\c_{\textstyle{.}}$'s
and $\fd$'s.
\eqn\fandchi{
\eqalign{
\c_{k_1,...,k_n}(\Gthd(X))&=
\sum_{I \cup J } (-1)^{ {\cal P}_{I,J}}
\left\{
\prod_{r<s \in I}
(P_{k_{r}}^{-1}+P_{k_{s}}^{-1}) \prod_{i \in I;
j \in J} (P_{k_i}^{-1}+P_{k_j}^{-1}) \right\} \times \cr
&\times
f_{k_{i_1},...,k_{i_{|I|}}}(X)
\delta_{k_{j_1},...,k_{j_{|J|}};|J|-1,|J|-2,...,0} \cr}}
Explicitly for $n=1,2,3$ this reads
\eqn\example{
\eqalign{
\c_k(X)&=p_k(X)=\delta_{k,0}+f_k(X) \cr
\c_{k_1,k_2}(X)&= \delta_{k_1,k_2;1,0}+f_{k_1-1}(X)
\delta_{k_2,0}+f_{k_1}(X) \delta_{k_2,1} \cr
&-f_{k_2-1}(X)
\delta_{k_1,0} -f_{k_2}(X) \delta_{k_1,1}
+f_{k_1-1,k_2}(X)+f_{k_1,k_2-1}(X) \cr
\c_{k_1,k_2,k_3}(X)&= \delta_{k_1,k_2,k_3;2,1,0} \cr
&+\sum_{{\rm cycl.}} ( f_{k_1-2}(X) \delta_{k_2,k_3;1,0}
+f_{k_1-1}(X) \delta_{k_2,k_3;2,0}+f_{k_1}(X)
\delta_{k_2,k_3;2,1} ) \cr
&+\sum_{{\rm cycl.}} \bigg([f_{k_1-1,k_2}(X) +
f_{k_1,k_2-1}(X)] \delta_{k_3,2} \cr
&+[f_{k_1-2,k_2-1}(X)+f_{k_1-1,k_2-2}(X)]\delta_{k_3,0}\cr
&+[f_{k_1-2,k_2}(X) +
2f_{k_1-1,k_2-1}(X)+f_{k_1,k_2-2}(X)] \delta_{k_3,1} \bigg) \cr
&+f_{k_1-2,k_2-1,k_3}(X)+f_{k_1-2,k_2,k_3-1}(X) \cr
&+
f_{k_1,k_2-2,k_3-1}(X)+f_{k_1-1,k_2-2,k_3}(X) \cr
&+f_{k_1-2,k_2,k_3-1}(X)+f_{k_1,k_2-1,k_3-2}(X)+
2 f_{k_1-1,k_2-1,k_3-1}(X). \cr}}
While $k_1,k_2,...,k_n$ are non negative on the l.h.s. of
\fandchi,
we recall once more that on the right hand side any $\fd$
with a negative or zero index is set equal to zero
\foot{ Eqn. \example\ suggests a
way of extending the definition of $f_{.}$'s to zero and
negative indices. For instance we could have defined
$\varphi_m=f_m$ for $m \geq 1$, and $\varphi_0=1$
to get $\c_k=\varphi_k=p_k$ for all $k \geq 0$.
In the more interesting case with two indices,
one can extend $\varphi_{k_1,k_2}=f_{k_1,k_2}$
for $k_1,k_2 \geq 1$, to $\varphi_{k_1,0}=f_{k_1}$ for
$k_1 \geq 1$, $\varphi_{0,k_2}=
-f_{k_2}$ for $k_2 \geq 1$, $\varphi_{0,0}=1$, and finally
$\varphi_{-1,1}=-2$, in order to get
$\c_{k_1,k_2}=\varphi_{k_1-1,k_2}+\varphi_{k_1,k_2-1}$
for all $k_1,k_2 \geq 0$.  For a general expression for
$\varphi$ see appendix A.}.

We can rewrite \fandchi\ as follows
\eqn\newfandchi{
\eqalign{
\c_{k_1+n-1,k_2+n-2,...,k_n}(X)&= \sum_{I \cup J}
(-1)^{{\cal P}_{I,J}}
\bigg\{ \prod_{r<s \in I} (1+ P_{k_r} P_{k_s}^{-1})
\prod_{i \in I ; j \in J} (1+P_{k_i} P_{k_j}^{-1}) \times \cr
&\times
\prod_{ p<q \in J } P_{k_p} \bigg\}
f_{k_{i_1},...,k_{i_{|I|}}}(X) \
\delta_{k_{j_1},...,k_{j_{|J|}};|J|-1,...,0} \cr} }
The operator $(1+Q)$, $Q=P_k P_l^{-1}$ admits as formal
inverse $(1+Q)^{-1}=\sum_{r \geq 0} (-Q)^r$
or $-\sum_{r \leq -1} (-Q)^r$.
When acting on both sides of \newfandchi, either form yield
equal finite sums as the reader will check.
We can therefore invert \newfandchi\ as
\eqn\invfandchi{
\eqalign{
\prod_{1 \leq i<j \leq n} (1 + P_{k_i} P_{k_j}^{-1})^{-1}
\c_{k_1+n-1,...,k_n}(X)&=
\sum_{I \cup J}  f_{k_{i_1},...,k_{i_{|I|}}}(X) \times \cr
&\times
\prod_{r < s \in J} P_{k_r} (1+P_{k_r} P_{k_s}^{-1})^{-1}
\delta_{k_{j_1},...,k_{j_{|J|}};|J|-1,...,0} \cr}}
For $k_1,...,k_n >0$ the only non-vanishing contribution
on the r.h.s. of \invfandchi\ corresponds to
$J=\emptyset$.
Indeed  when $J \ne \emptyset$,
the antisymmetric Kronecker
symbols always contain the constraint that at least one
of the $k_i$ be zero or negative. Consequently, with $\bf r$
an antisymmetric  $\IZ$-valued $n \times n$ matrix,
if we mean by ${\bf r} \geq 0$ the conditions $r_{ij}\geq 0$
for $i<j$, we find the inversion formula
\eqn\fininv{
\encadremath{
f_{k_1,...,k_n}(X)=\sum_{{\bf r} \geq 0} (-1)^{\GS_{i<j}r_{ij}}
\c_{k_1+n-1+\GS_j r_{1j},k_2+n-2+\GS_j r_{2j},...,
k_n+\GS_j r_{nj}}(X) }}
The sums over the $r$'s on the r.h.s. are finite, since $\c_{\textstyle{.}}(X)$
vanishes whenever $k_p+n-p+\sum_j r_{pj} <0$.
We see that $\fd$ is indeed a polynomial in $\Gthd(X)$
and since no reference is made to the size of the matrix $X$,
the two inverse formulas \fandchi\-\invfandchi\ can be
thought of as relating two families of polynomials in
infinitely many variables $\theta_1,\theta_3,...$.
Furthermore,
\eqn\fdegree{ \deg f_{k_1,...,k_n}(\Gthd)= \sum_{s=1}^n k_s }
we have therefore

\noindent{ \bf Lemma 2} Any polynomial in $\theta_1,\theta_3,...$
admits a unique expansion in terms of $f \equiv 1$,
$f_k ,k>0$, $f_{k_1,k_2},k_1>k_2>0$,...

Uniqueness is a consequence of a dimensional argument.
The dimension of the vector space of polynomials of degree $n>0$
in $\theta_1,\theta_3,...$ is equal to the number of partitions of $n$
in odd integers, while the dimension of the linear span of $\fd$'s
such that $k_1>k_2>...>k_r>0$ $\sum_{1 \leq s \leq r} k_s = n$,
$r>0$, is the number of partitions of $n$ into unequal parts.
The two are equal by virtue of Euler's identity
\eqn\euler{
\eqalign{
\prod_{n>0} (1+q^n)&= \prod_{n \geq 0} {1 \over {(1-q^{2n+1})}} \cr
&=
1+q+q^2+2q^3+2q^4+3q^5+4q^6+5q^7+6q^8+... \cr}}
and of course all coefficients in $\fd$'s are again rational.

The reader will find in appendix A some relations
enabling to compute $\fd$'s efficiently. We illustrate the
change of basis from $\fd$'s to monomials in $\Gthd$
in tables I and II, where we use the notation
$\theta_{[1^{\nu_1} 3^{\nu_3} ...]}$ for the monomial
${\theta_1^{\nu_1} \over \nu_1!} {\theta_3^{\nu_3}
\over \nu_3!}...$.

\subsec{Averages}

To complete the proof of (K') it is now sufficient to perform the
averages $\langle \fd(\Gthd(Y)) \rangle_{(N)}$.
For this we insert the original definition \fdef, taking the size of the
matrices large enough (in particular
$N \ge n$, the number of indices).
Of course $n>0$, since for $n=0$ we have nothing to prove.
Thus
\eqn\averages{
\eqalign{
\langle f_{k_1,...,k_n}(\Gthd(Y)) \rangle_{(N)}&=
{{(-1)^{N(N-1)/2} 2^n (-1)^{k_1+...+k_n} } \over
{ N! Z_N(\gL) }} \int \prod_{a=1}^n {{d y_a} \over \sqrt{ 2 \pi}}
\det [ e^{-{{y_a^2 \gl_b} \over 2}} ]_{1 \leq a,b \leq N}
\times \cr
&\times \prod_{1 \leq a < b \leq N} {{y_a-y_b} \over {(\gl_a-\gl_b)
(y_a+y_b)}}
\sum_{1\leq a_1<...<a_n \leq N} \prod_{1 \leq r<s \leq n}
{{y_{a_r}+y_{a_s}} \over {y_{a_r}-y_{a_s}}}  \times \cr
&\times
\prod_{1 \leq t \leq n \atop l \in \{1,..,{\hat a_1},...,{\hat a_l},..,N\}}
{{y_{a_t}+y_l} \over {y_{a_t}-y_l}} \ \
\det[ y_{a_m}^{k_p}]_{1 \leq m,p \leq n} \cr}}
We expand the $N \times N$ determinant
$\det e^{-{{y_a^2 \gl_b}\over 2}}$ according to Lagrange's
formula as an alternating sum of products of determinants
of size $n$ and $N-n$ respectively, take signs carefully into
account and note vast cancellations, to get
\eqn\tiensbon{
\eqalign{
\langle f_{k_1,...,k_n}(\Gthd(Y))\rangle_{(N)}&=
{{(-1)^{N(N-1)/2} 2^n (-1)^{k_1+...+k_n} } \over
{ N! Z_N(\gL) \prod_{1\leq a<b \leq N} (\gl_a -\gl_b)}}
\int \prod_{a=1}^n {{d y_a} \over \sqrt{ 2 \pi}}
\sum_{1 \leq a_1<...<a_n \leq N \atop
1 \leq b_1<...<b_n \leq N} \times \cr
&\prod_{a<a'\atop
a,a' \in \{1,..,{\hat a_1},..,{\hat a_n},..,N\}}
(-1)^{b_1+..+b_n}
{{y_a-y_{a'}} \over {y_a+y_{a'}}}
\det[ e^{-{{y_a^2 \gl_b} \over 2}}]_{
a \in \{1,..,{\hat a_1},..,{\hat a_n},..N\} \atop
b \in \{1,..,{\hat b_1},..,{\hat b_n},..,N\}} \times \cr
&\times \det[e^{-{{y_{a_s}^2 \gl_{b_t}} \over 2}}]_{
1 \leq s,t \leq n} \det[y_{a_s}^{k_t}]_{1 \leq s,t \leq n}
\cr } }
For each set $1 \leq a_1<a_2..<a_n \leq N$ the integral
over the corresponding $y$'s yields equal results while
the integral over the remaining $y$'s yields
a factor $Z_{N-n}(\gL_{b_1,..,b_n})$
\eqn\galere{
\eqalign{
\langle f_{k_1,...,k_n}(\Gthd(Y)) \rangle_{(N)}&=
{{(-1)^{n(2N-n-1)/2} 2^n (-1)^{k_1+..+k_n} }
\over { n!}} \sum_{1 \leq b_1<..<b_n \leq N}
{{Z_{N-n}(\gL_{b_1,..,b_n})} \over Z_{N}(\gL)} \times \cr
&\times (-1)^{b_1+..+b_n}
{{\prod_{b<b' \in \{1,..,{\hat b_1},..,{\hat b_n},..,N\}}
(\gl_b -\gl_{b'}) } \over {\prod_{1 \leq b<b' \leq n}
(\gl_b-\gl_{b'})}} \times \cr
&\times \int
\prod_{t=1}^n {{d y_t} \over \sqrt{2 \pi}}
\det [e^{-{{y_s^2 b_t} \over 2}}]_{1 \leq s,t \leq n}
\det [y_s^{k_t}]_{1 \leq s,t \leq n} \cr }}
The last integral vanishes whenever one of the
$k_i$ at least is odd, hence we get \main a\
\eqn\profmaina{ \langle f_{k_1,...,k_n}(\Gthd(Y))
\rangle_{(N)} = 0 \ \ \ {\rm if \ at \ least \ one \ } k_i \ne 0
\ {\rm mod} \ 2}
whereas using $\int dy\, y^{2k} e^{-{\gl y^2/2}}=\sqrt{{2 \pi} / \gl}
(2k-1)!!  \gl^{-k} $, we get
\eqn\profmainb{
\eqalign{
\langle f_{2k_1,...,2k_n}&(\Gthd(Y)) \rangle_{(N)}=
2^n (-1)^{n(2N-n-1)/2} \prod_{s=1}^n (2k_s-1)!! \times \cr
&\times \sum_{1 \leq b_1<..<b_n \leq N}
\prod_{1 \leq s<t \leq n} {{\gl_{b_s} + \gl_{b_t}} \over
{\gl_{b_s} -\gl_{b_t}}}
\prod_{1 \leq s \leq n \atop
b \in \{ 1,..,{\hat b_1},..,{\hat b_n},..,N\}}
{{\gl_{{b_s}}+ \gl_{b}} \over {\gl_{b_s} - \gl_{b}}}
\det[\gl_{b_t}^{k_s}]_{1 \leq s,t \leq n} \cr }}
Since $(\gl_b + \gl_{b'})/(\gl_b-\gl_{b'})=
-(\gl_b^{-1}+\gl_{b'}^{-1})/(\gl_b^{-1} - \gl_{b'}^{-1})$,
comparing with the definition of $\fd(\Gthd(\gL^{-1}))$,
we obtain \main b\ in the form
\eqn\mbproof{
\langle f_{2k_1,...,2k_n}(\Gthd(Y)) \rangle_{(N)}=
(-1)^{k_1+..+k_n} \prod_{s=1}^n (2k_s-1)!! \
f_{k_1,...,k_n}(\Gthd(\gL^{-1})) }
as claimed. This completes the proof of proposition (K').
We add a few comments.

(i) As shown along the way the map $K$: $P(\Gthd)
\to Q(\Gthd)$ is defined over $\IQ$.

(ii) Since $\fd$'s generate all polynomials in odd traces,
this map is surjective. Its kernel is the linear span
$${\rm Ker} (K)=\{ {\rm linear \ span \ of \ }
f_{k_1,...,k_n}(\Gthd), \ n>0, {\rm\ such \ that \ at
\ least \ one \ of \ the \ }
k_i  {\rm \ is \  odd} \} $$
Indeed any $P(\Gthd)$ is a finite sum
\eqn\pfinsum{ P(\Gthd)= a_0+ \sum_{n>0}
\sum_{k_1>k_2>...>k_n>0} a_{k_1,..,k_n} \
f_{k_1,...,k_n}(\Gthd) }
if $K(P)=0$, it follows that
\eqn\follo{
\sum_{k_1>...>k_n>0} a_{2k_1,..,2k_n}\
f_{k_1,..,k_n}(\Gthd) =0 \Rightarrow a_{2k_1,..,2k_n}=0}
Therefore if $d(n)$ denotes the dimension of the vector space
of polynomials of degree $n$ in $\Gthd$'s and $d_0(n)$ the
dimension
of the subspace annihilated by $K$, we have
\eqn\count{
\eqalign{
\sum_{n=0}^{\infty} d(n) q^n &= \prod_{n>0}(1 + q^n) \cr
\sum_{n=0}^{\infty} d_0(n) q^n&= \prod_{n>0}(1+q^n)
-\prod_{n>0} (1+q^{2n}) \cr
&=\prod_{n \geq 0} {1 \over {(1-q^{2n+1})}} \left( 1-
\prod_{n \geq 0} { 1 \over { (1+q^{2n+1})}} \right) \cr
&= q+ 2q^3 + q^4+3q^5+2q^6+5q^7+4q^8+... \cr}}

\newsec{Proof of Proposition (W)}

\noindent Notations being as before, we consider the integral \kont\
\eqn\kontiki{ \Xi_{N}(\gL^{-1})= {{ \int \d \mu_{\gL}^{(N)}(Y)
\exp{{i \over 6} \tr Y^3} } \over { \int \d \mu_{\gL}^{(N)}(Y) }}
= \langle \exp{{i \over 6} \tr Y^3 } \rangle_{(N)}(\gL^{-1}) }
which admits an asymptotic expansion, each term
of which is for $N$ sufficiently large an $N$-independent
polynomial in the odd traces of $\Lambda^{-1}$.
We keep the normalization \northeta\
\eqn\nornou{ \theta_{2k+1}(\Lambda^{-1})=-{2 \over {2k+1}}
{\rm Tr} \Lambda^{-2n-1} }
As $N$ tends to infinity these become independent variables
and the asymptotic expansion is denoted $\Xi(\Gthd)$.
For any polynomial in odd traces $P(Y) \equiv P({\rm tr }Y,
{\rm tr}Y^3,...)$ set
\eqn\sset{ \ll P \gg_N = \langle P(Y) \exp({i \over 6}
{\rm tr}Y^3 \rangle_N(\Lambda^{-1}) }
{}From section 2 it admits an $N$-independent asymptotic expansion
$\ll P \gg(\Gthd)$ in the odd traces $\theta_{2k+1}(\Lambda^{-1})$.
Explicit calculations performed by Witten \WIT\ suggest
that one can express this average as
\eqn\pliwit{ \ll P \gg = R({\partial \over \partial {\Gthd}})
\Xi(\Gthd) }
where $R$ is again a polynomial (with constant coefficients) in the
derivatives $\partial \over \partial {\theta_{2k+1}}$, and
conversely that for any such $R$ there exists a $P$.
Using techniques developed in \CJB, one can for instance derive
closed expressions for $R$ a monomial in $\theta_1$ or $\theta_3$
(see appendix B)
\eqn\iandiii{
\eqalign{
\( {\partial \over \partial {\theta_1} }\)^k \Xi(\Gthd)
&= \sum_{m,n \geq 0 \atop 3n+m = k} {{(3n+m)!} \over{
6^n n! m!}} \ll \( {\rm tr} {Y \over 2i} \)^m \gg \cr
\( {\partial \over \partial {\theta_3}} \)^k \Xi(\Gthd) &=
\( (1+3y/4)^3 \partial_y \)^k
(1+3y/4)^{1 \over 12} \ll e^{y {\rm tr} \big( {Y \over 2i} \big)^3}
\gg \vert_{y=0}  \cr }}
and a more cumbersome, although perfectly explicit formula
for $R$ being  any polynomial in both $\partial_{\theta_1}$ and
$\partial_{\theta_2}$
(see appendix B for details).

The proof of proposition (W) is based as before on the comparison
of the integral over $N \times N$
matrices $\Xi_{N}$,  defined in \kontiki, to
the same integral $\Xi_{N+n}$ over $(N+n) \times(N+n)$
matrices.
Indeed it is easy to see for $n=1$ that  in the expansion of
$\Xi_{N+1}(\gl^{-1}\oplus\Lambda^{-1})$, where $\lambda$
is a real positive number and $\Lambda$ a positive definite
diagonal $N \times N$ matrix, the terms
of degree
$3k \leq N$ are obtained
from those of the same degree
in the expansion of
$\Xi_N(\Lambda^{-1})$ by translating
the variables $\Gthd(\Lambda^{-1})$ according to
\eqn\trans{ \theta_{2j+1}(\Lambda^{-1}) \to
\theta_{2j+1}(\gl^{-1} \oplus \Lambda^{-1})=
\theta_{2j+1}(\Lambda^{-1}) -{2 \over 2j+1} \gl^{-2j-1}}
Therefore in the usual $N \to \infty$ limit, we can write
\def\ddd{\partial_{\textstyle{.}}}
\eqn\transchur{
\eqalign{
\Xi(\gl^{-1}\oplus\Lambda^{-1})
&= \exp{{\scriptstyle\sum}_{i=0}^{\infty} -{2 \over 2j+1} \gl^{-2j-1}
{\partial \over \partial {\theta_{2j+1}}}} \
\Xi(\Lambda^{-1}) \cr
&= \sum_{k=0}^{\infty} \gl^{-k} \ p_k(\ddd) \ \Xi \cr}}
We see that the power series in $\gl^{-1}$ on the r.h.s. is a
generating function for the Schur polynomials (defined
in \schur) of derivatives $\ddd$
\eqn\defder{
\eqalign{
\partial_{2j+1} &\equiv -{2 \over {2j+1}}
{\partial \over \partial {\theta_{2j+1}}} \cr
\partial_{2j} &\equiv  0 \cr}}
acting on $\Xi$, as a function of the infinitely many variables
$\theta_{2j+1}$.
Increasing $n$ amounts to iterating this process, and
we get in general a generating function for
any product of Schur polynomials of $\ddd$
acting on $\Xi$.  These products span the whole space of
polynomials in the variables $\partial_{2j+1}$.

We are left with the task of computing the l.h.s. of
\transchur\ and its generalizations.
Proposition (W) will follow if we can find expressions of the former
as generating functions for
expectation values of polynomials of the form \avkon.
This last step turns out to be elementary, and leads
to explicit expressions for the
aforementioned polynomials. It is summarized in the following

\medskip

\noindent{\bf Lemma 3}
Let $\Lambda=\Lambda_1 \oplus \Lambda_2$
be the decomposition of the diagonal matrix
$\Lambda={\rm diag}(\gl_1,...,\gl_{N+n})$ into the direct
sum
of two diagonal matrices $\Lambda_1={\rm diag}(\gl_1,...,
\gl_n)$ and
$\Lambda_2={\rm diag}(\gl_{n+1},..,\gl_{n+N})$. We have
\eqn\reswpr{
\encadremath{
\eqalign{
\prod_{1\leq i<j \leq n} &{{\gl_i -\gl_j} \over{\gl_i + \gl_j}}\
\Xi_{n+N}(\Lambda_1^{-1} \oplus \Lambda_2^{-1})=
\int \prod_{k=1}^n  d\nu_{\gl_k}(y_k) \times \cr
&\prod_{1 \leq m<p \leq n} {{2i(\gl_m -\gl_p)+y_m - y_p}\over
{2i(\gl_m + \gl_p) + y_m + y_p}}\
\ll \prod_{l=1}^n \det \(
{{2i \gl_l + y_l -Y_2} \over {2i \gl_l + y_l + Y_2}} \) \gg_N
(\Lambda_2^{-1})  \cr}}}
where $d \nu_{\gl}(y)= (\gl/2\pi)^{1 \over 2}
\exp(iy^3/6 -\gl y^2/2) dy $ is the measure  of integration
over the eigenvalues $y$ adapted to our problem, and
the double bracket denotes the integral over the
$N \times N$
matrix $Y_2$ as defined in \sset.

\medskip
The lemma will be used to expand both sides of \reswpr, when
$N \to \infty$,
as formal series in $\gl_j^{-1}$, $1 \leq j \leq n$, for large $\gl$'s.
On the l.h.s. of \reswpr\ we get as coefficients of this series
polynomials in the derivatives w.r.t. $\Gthd(\Lambda_2^{-1})$,
whereas on the r.h.s. one gets averages of polynomials in the
odd traces of $Y_2$, which completes the proof of proposition (W).
Before proving lemma 3, let us illustrate the mechanism
in the case $n=1$, where \reswpr\ reduces to
\eqn\resiw{ \Xi_{N+1}(\gl^{-1} \oplus \Lambda^{-1})=
\int d\nu_{\gl}(y) \ll \det\( {{ y+2i \gl - Y} \over { y+2i \gl + Y}} \) \gg_N
}
Expanding both sides in powers of $\gl^{-1}$, using
\eqn\perexn{ \Xi_{N+1}(\gl^{-1} \oplus \gL^{-1})=
\int d \nu_{\gl}(y) \ll \sum_{m \geq 0} (\gl - {iy \over 2})^{-m}
p_m \big(\Gthd({Y \over 2i}) \big) \gg_N }
with $N \to \infty$,
and integrating term by term over $y$,
we can identify the coefficient of
$\gl^{-k}$ in \transchur\ as
\eqn\identschur{
\encadremath{ p_k(\ddd) \ \Xi (\Gthd)=
\sum_{0 \leq s \leq [k/3]} (-1)^s  c_{s,k}  \ll p_{k-3s}\big(\Gthd({ Y \over
2i})
\big) \gg_N}}
where
\eqn\coefcsk{ c_{s,k} = \sum_{l=0}^{2s} {1 \over 2^l} { {(k-3s+l-1)!} \over
{l! (k-3s-1)!}} {{(6s-2l-1)!!} \over {6^{2s-l} (2s-l)!}} }
and $[x]$ denotes the integral part of $x$. For $k=0$, \identschur\
reduces to the identity $\ll 1 \gg =\Xi(\Gthd)$.
The general case with $n>1$ will be dealt with below.
Let us turn to the proof of lemma 3.

At first the matrices $\Lambda,\Lambda_1,\Lambda_2$ involve diagonal real
positive elements, but if we introduce a cut in the complex plane along
the negative real axis, the integrals make sense for each eigenvalue having a
positive real part -- as absolutely convergent integrals; as semi-convergent
ones
we can even extend them to the imaginary axis except the origin.
To give a meaning to the following operations we will first continue
analytically the $\gl_j$ to imaginary non vanishing values.
Similar techniques were implicit in both \KON\ and \CJB.
With this proviso in mind
we return to
\eqn\startex{ \Xi_{n+N}(\Lambda^{-1})=
{ 1 \over Z_{n+N}(\Lambda)} \int dY e^{{i \over 6}{\rm tr}(Y^3)
-{1 \over 2} {\rm tr}(\Lambda Y^2)}}
where $Z_{n+N}(\Lambda)$ is defined in \gauss.
We perform the change of variables $Z=Y+2i (\Lambda_1\oplus 0)$,
with the obvious definition for the $(n+N)\times
(n+N)$ matrix $\Lambda_1 \oplus 0 ={\rm diag}(\gl_1,..,\gl_n,0,..,0)$.
Due to the relation
$$ (\Lambda_1 \oplus 0)(0 \oplus \Lambda_2)=
(0 \oplus \Lambda_2)(\Lambda_1 \oplus 0)=0 $$
the trace in the exponential becomes
\eqn\tracex{ {i \over 6}{\rm tr}(Y^3) -
{1 \over 2}{\rm tr}(\Lambda Y^2)= {i \over 6}{\rm tr}(Z^3)
-{1 \over 2} {\rm tr}( [(0 \oplus \Lambda_2) - (\Lambda_1 \oplus 0)]Z^2)
+{2 \over 3} {\rm tr}(\Lambda_1^3) }
We see that except for a constant term, the form of the
exponential term is conserved, up to the substitution
$\Lambda=\Lambda_1\oplus\Lambda_2 \to { \tilde \Lambda}=
(0 \oplus \Lambda_2)-(\Lambda_1 \oplus 0)$. We can now perform the
``angular" average over $Z$ \INT, which results in
\eqn\intermede{ \Xi_{n+N}(\Lambda^{-1})=\prod_{1 \leq i<j \leq N+n}
{{ \gl_j + \gl_i} \over {{\tilde {\gl_j}} - { \tilde { \gl_i}}}}
\int \prod_{k=1}^{n+N}  d \nu_{\tilde {\gl_k}}(z_k)
\prod_{p=1}^n e^{{2 \over 3} \gl_p^3}
\prod_{1 \leq l<m \leq N+n} {{ z_l - z_m} \over {z_l+ z_m}} }
where the ${\tilde {\gl}}$'s are the diagonal elements of
${\tilde {\Lambda}}$, i.e. ${\tilde {\gl_k}}= -\gl_k$ for
$1 \leq k \leq n$, ${\tilde {\gl_k}}=\gl_k$ for $n+1 \leq k \leq n+N$
(recall that the $\tilde \gl$'s are purely imaginary, so that the
minus sign causes no harm in the integral).
As usual the antisymmetry of the integrand in $z$'s in \intermede\
automatically takes care of the denominators $z_l+z_m$,
by antisymmetrizing the measure.
We now perform the opposite change of variables, but this time on the
eigenvalues $z$ by setting
\eqn\invchgvar{
\eqalign{
z_k&= y_k + 2i \gl_k  \qquad  1\leq k \leq n \cr
z_k &= y_k  \ \ \   \qquad  n+1 \leq k \leq n+N \cr}}
which leads to
\eqn\finlem{
\eqalign{
\Xi_{n+N}&(\Lambda_1^{-1} \oplus \Lambda_2^{-1}) =
\prod_{1 \leq i<j \leq n} {{ \gl_i + \gl_j} \over { \gl_i -\gl_j}}
\prod_{n+1 \leq l<m \leq n+N} {{ \gl_l^{-1} + \gl_m^{-1}}
\over { \gl_l^{-1} - \gl_m^{-1}} } \int
\prod_{k=1}^{N+n} d \nu_{\gl_k}(y_k) \times \cr
&\times \prod_{1 \leq l<m \leq n} {{y_l -y_m+2i(\gl_l -\gl_m)}
\over{ y_l+y_m + 2i(\gl_l + \gl_m)}}
\prod_{n+1 \leq i<j \leq n+N} {{ y_i - y_j} \over {y_i+y_j}}
\prod_{1 \leq l \leq n \atop n+1 \leq j \leq n+N}
{{ y_l + 2i \lambda_l - y_j} \over { y_l + 2i \gl_l + y_j}}  \cr }}
and amounts to the statement of lemma 3 since
\eqn\finlemiii{
\eqalign{
\prod_{n+1 \leq l<m \leq n+N}
&{{\gl_l^{-1} + \gl_m^{-1}} \over {\gl_l^{-1} -\gl_m^{-1}}}
\int \prod_{k=n+1}^{n+N} d \nu_{\gl_k}(y_k)
\prod_{n+1 \leq i<j< n+N} {{ y_i -y_j} \over {y_i + y_j}} \times \cr
&\times \prod_{1 \leq l \leq n \atop n+1 \leq j \leq n+N}
{{ y_l + 2i \gl_l -y_j} \over {y_l + 2i \gl_l +y_j}}
= \ll \prod_{1 \leq l \leq n}
\det \( {{ y_l + 2i \gl_l -Y_2} \over { y_l + 2i \gl_l + Y_2}} \)
\gg_N(\Lambda_2^{-1}) \cr}}
\noindent{\bf Remarks}\item{(i)}Once we are through with the
proof, the lemma remains valid as a statement on asymptotic series
for real positive eigenvalues $\gl$'s.
\item{(ii)}The lemma enables us to give several
expressions for the same $\Xi_N$ by splitting the $N$
eigenvalues of $\Lambda$  into two sets $N=N_1 + N_2$.
For instance we find for $N=2$ the various expressions
with $\Lambda={\rm diag} (\gl_1,\gl_2)$
\noindent{}
\eqna\twocase
$$\eqalignno{
\Xi_2(\Lambda^{-1}) &= \int \prod_{k=1,2} d \nu_{\gl_k}(y_k)
{{ y_1 -y_2 + 2i( \gl_1 -\gl_2) } \over { y_1+y_2+2i(\gl_1+\gl_2)}}
\times {{ \gl_1 + \gl_2 } \over { \gl_1 -\gl_2}}
&\twocase a \cr
&= \int \prod_{k=1,2} d \nu_{\gl_k}(y_k)
{{ y_1 + 2i \gl_1 - y_2} \over { y_1+2i \gl_1 + y_2}}
&\twocase b \cr
&= \int \prod_{k=1,2} d \nu_{\gl_k}(y_k)
{{ y_1 - y_2 } \over {y_1+y_2}}
\times {{\gl_1+ \gl_2} \over {\gl_2 - \gl_1}}
&\twocase c \cr }$$
where \twocase b\ admits an alternative expression with
$\gl_1 \leftrightarrow \gl_2$ and \twocase c\
is understood after antisymmetrization of the measure.
It is interesting to check that the difference of any two of these
expressions can be written as the integral of a total derivative
(and therefore vanishes)!

With lemma 3 in hand, we can now obtain the most general
combination of derivatives acting on $\Xi(\Gthd)$.
The l.h.s. of \reswpr\ for $N \to \infty$
converges as an asymptotic series
in $\Gthd(\Lambda^{-1})$ to
\eqn\thefscome{
\eqalign{
\prod_{1 \leq i<j \leq n} {{\gl_i -\gl_j} \over {\gl_i+\gl_j}} \
&\Xi(\gl_1^{-1} \oplus ..\oplus \gl_n^{-1} \oplus
\Lambda^{-1})  \cr
&=\prod_{1 \leq i<j \leq n}
{{\gl_i -\gl_j} \over {\gl_i+\gl_j}}
\sum_{r_1,r_2,..,r_n} \prod_{i=1}^n \gl_i^{-r_i}
p_{r_1}(\ddd) ... p_{r_n}(\ddd)
\ \Xi(\Gthd(\Lambda^{-1})) \cr}}
Expanding the prefactor as a power series in the
domain $\gl_1^{-1} > \gl_2^{-1}>...>\gl_n^{-1}$ and
performing a similar expansion on the r.h.s. of \reswpr,
it is possible to rearrange the series in terms of our $\fd$'s.
More precisely, it is simpler to use
the extension $\varphi$ of $f$ described in appendix A,
to zero or negative indices,
generated by the series
\eqn\genfext{
\eqalign{
F_n(\gl_1^{-1},..,\gl_n^{-1};X)&=
\sum_{m_i \in {\IZ}} \prod_{k=1}^n \gl_k^{-m_k}
\varphi_{m_1,..m_n}(\Gthd(X)) \cr
&= \prod_{1 \leq i<j \leq n} {{ \gl_i^{-1} -\gl_j^{-1}} \over
{\gl_i^{-1} + \gl_j^{-1} } }  \sum_{n_i \geq 0} \prod_{i=1}^n
\gl_i^{-n_i} \ p_{n_i}(\Gthd(X)) \cr} }
expanded in the domain $\gl_1^{-1} > .... > \gl_n^{-1}$.
The l.h.s. of \reswpr\ reads therefore
\eqn\phicomes{
\eqalign{
(-1)^{{n(n-1) \over 2}} \prod_{1 \leq i<j \leq n}
{{\gl_i^{-1} - \gl_j^{-1} } \over { \gl_i^{-1} + \gl_j^{-1}}}
\ &\Xi(\gl_1^{-1}\oplus..\oplus \gl_n^{-1}\oplus
\Lambda^{-1})  \cr
&= (-1)^{n(n-1) \over 2}
\sum_{m_k \in {\IZ} } \prod_{k=1}^{n}
\gl_k^{-m_k}  \ \varphi_{m_1,...,m_n}(\ddd)
\ \Xi(\Lambda^{-1}) \cr}}
\def\phid{\varphi_{{\textstyle .}}}
To perform such an  expansion on the r.h.s. of \reswpr\
we should  similarly order the
arguments $\gl_k+(y_k/2i)$ (in modulus).
This looks at first unreasonable since the $y$'s run along the
whole real axis.
However we recall that we look for an asymptotic
expansion of an absolutely
convergent integral over $y_1,..,y_n$, as each $\gl_1,..,\gl_n$
goes to infinity.
With an exponentially small error we can therefore bound the
domain of integration in the $y$'s and assume the $\gl$'s
large enough so that the $|\gl_k+(y_k/2i)|$ remain ordered.
This means that in the sense of {\it asymptotic series}
we can perform an expansion similar to the above \phicomes\
but this time with  $\phid(Y/2i)$ in the integrand.
Integrating on the $y$'s as we did before, we can identify the
coefficient of $\gl_1^{-m_1}...\gl_n^{-m_n}$,
$m_1>...>m_n>0$ in the resulting asymptotic expansion.
We obtain the main result of this section in the form of

\medskip
\noindent{\bf Proposition (W')} Any polynomial in derivatives acting
on $\Xi(\Gthd)$ can be expressed as an average over a polynomial in
odd traces of $Y$ and vice versa. More precisely for $m_1>..>m_n>0$
\eqn\mapq{
\encadremath{
f_{m_1,..,m_n}(\ddd)\ \Xi(\Gthd)=
\sum_{s_1,..,s_n \geq 0  \atop 3s_j \leq\sum_{j \leq k \leq n} m_k}
\prod_{i=1}^n (-1)^{s_i} c_{s_i,m_i} \ll
\varphi_{m_1-3s_1,...,m_n-3s_n}(\Gthd(Y/2i)) \gg_N }}

\medskip
We recall the notations and make a few comments.
\item{(a)} $\ddd= \{-{2 \over {2k+1}}
{\partial \over {\partial \theta_{2k+1}}} \}$.
\item{(b)} the $\phid$'s are reduced to $\fd$'s with
positive indices according to the rules (i)-(iv) of appendix A.
\item{(c)} $c_{s,m}=\sum_{l=0}^{2s} {1 \over 2^l}
{m-3s+l-1 \choose l} {{(6s-2l-1)!!} \over {6^{2s-l} (2s-l)!}} $, where the
combinatorial factor can be seen as a polynomial in the variable
$m-3s$ with integral coefficients, hence remains
integral and well defined
for $m-3s<0$.
It is easy to rewrite
$$ c_{s,m}={1 \over (12)^{2s}} \sum_{l=0}^{2s} 6^l  {{(6s-2l-1)!!}
\over {(4s-2l-1)!!}} {4s-2l \choose 2s-l}{m-3s+l-1 \choose l} $$
exhibiting $(12)^{2s} c_{s,m}$ as an integer.
Since the $\phid$'s are linear combinations of $\fd$'s
with integral coefficients (see appendix A), the relation
\mapq\ involves at most rational fractions
with denominators $(12)^{2s_i}$ as coefficients.
\item{(d)} The ``leading" term in \mapq\ corresponds to $s_1=...=s_n=0$
and $c_{o,m}=1$, hence for $m_1>...>m_n>0$
\eqn\leading{ f_{m_1,..,m_n}(\ddd) \ \Xi(\Gthd)=
\ll f_{m_1,..,m_n}(\Gthd(Y/2i)) \gg_N +\  {\rm lower} ...}
where by lower we mean averages over polynomials with smaller degree
in $Y$. Therefore the system of equations \mapq\ is triangular and can
be inverted, vindicating the statement (W').
\item{(e)} In table III we have recorded explicitely the first few cases
of \mapq\ up to degree $8$ (the argument in the average is
$\Gthd(Y/2i)$). The reader can check --as we did in table IV--
that these data are in agreement with formulas \iandiii\
of this section, expressing derivatives w.r.t.
$\theta_1$ and $\theta_3$. Also one can make contact with
earlier results by E. Witten \WIT\ expressed in terms of
variables $t_{.}$ related to our $\Gthd$'s \nort\ through
\eqn\tthet{ t_{k} ={(2k+1)!! \over 2} \theta_{2k+1} }
\item{(f)} There exist simple
cases of \mapq\ when successive $m$'s differ by $3$ and the
last one is $1$ or $2$, where, due to the antisymmetry of $\fd$'s,
the relation \mapq\ reduces to
\eqn\remfin{ f_{a+3k,a+3(k-1),..,a+3,a}(\ddd) \ \Xi(\Gthd)=
\ll f_{a+3k,a+3(k-1),...,a+3,a}(\Gthd(Y/2i)) \gg_N }
for $a=1,2$.


\vfill
\eject

%
%

\appendix {A} {Calculation of ${\bf f}_{\scriptstyle{\bf .}}$'s}

To make the computation of the polynomials
$\fd$ simple and explicit,
let us use the definition of the characters $\c_{\textstyle .}$ in terms
of our Schur polynomials $p_{.}$ \schur\ to rewrite \fininv\ as
\eqn\plicit{
\eqalign{
f_{k_1,....,k_n}(\Gthd)&= \left\{\prod_{1 \leq i < j \leq n}
{{1-P_{k_i} P_{k_j}^{-1}} \over {1 + P_{k_i} P_{k_j}^{-1}}} \right\}
p_{k_1}(\Gthd)
p_{k_2}(\Gthd)...p_{k_n}(\Gthd) \cr
&= \sum_{{\bf r} \geq 0} \(
\prod_{1 \leq i<j \leq n} \alpha_{r_{ij}} \)
p_{k_1+\GS_j r_{1j}}(\Gthd) ...
p_{k_n+\GS_j r_{nj}}(\Gthd) \cr}}
where $\alpha_r=(-1)^r(2-\delta_{r,0})$ is the coefficient of $y^r$
in the small $y$ expansion of $(1-y)/(1+y)$, and the  sum
over $\bf r$ (defined as in \fininv) is finite ($k_p+\sum_j r_{pj} \geq 0$).
The second line of \plicit\ makes it straightforward to extend the
definition of $\fd$'s to $\phid$'s including negative or zero indices.
Namely define for $m_1,..,m_n \in {\IZ}$
\eqn\phienfin{
\varphi_{m_1,..,m_n} = \sum_{ {\bf r} \geq 0}
\(\prod_{1 \leq i<j \leq n} \alpha_{r_{ij}} \)p_{m_1+\sum_i r_{1i}}...
p_{m_n+\sum_i r_{ni}} }
with the convention that $p_m=0$ as soon as $m<0$, then one has obviously
\eqn\idphif{ \varphi_{m_1,..,m_n}=
f_{m_1,...,m_n} \ \ \ { \rm for} \ m_1,...,m_n>0}
This definition enables one to rewrite the expressions \fandchi\-\example\
for the characters in a very simple way
\eqn\charplus{
\c_{\textstyle{.}}=\prod_{i<j} (P_i^{-1}+ P_j^{-1}) \phid}
The generating function \compchif\ for characters also simplifies
drastically. Namely in the region $|u_1|>|u_2|>...>|u_n|$ of the
polydisk $D(X)^n$, we can expand $F$ defined in \genfunc\ as
\eqn\genphi{
\eqalign{
F_n(u_1,...,u_n;X)&= \prod_{1 \leq i<j \leq n} {{ 1-(u_j/u_i)} \over
{1+(u_j / u_i)}} \prod_{k=1}^n \det {{1-u_k X} \over {1+u_k X}} \cr
&=\sum_{{\bf r} \geq 0}  \sum_{k_1,...,k_n \geq 0}
\prod_{1 \leq i<j \leq n} \alpha_{r_{ij}} (u_j/u_i)^{r_{ij}}
\prod_{m=1}^n u_m^{k_m}\  p_{k_m+\sum_j r_{mj}}(X) \cr
&= \sum_{m_1,..,m_n \in {\IZ}} \(\prod_{i=1}^n u_i^{m_i} \)
\ \varphi_{m_1,...,m_n}(X) \cr}}

But as shown in section 2, the $\fd$'s form a basis of the
vector space of polynomials in the variable $\Gthd$, therefore
the extension $\phid$ is overcomplete. In fact as already apparent
on the r.h.s. of \compchif, the $\phid$'s can be expressed
simply in terms of $\fd$'s only: we get the following set of rules
easily derived from the definition \phienfin\
\item{(i)} $\varphi_{m_1,...,m_n}=f_{m_1,...,m_n}$, for all $m_1>0$.
\item{(ii)} $\varphi_{m_1,...,m_n,0}=\varphi_{m_1,...,m_n}$, for all $m_i \in
\IZ$.
\item{(iii)} $\varphi_{m_1,...,m_n,a}=0 $, whenever $a<0$.
\item{(iv)}  $\varphi_{m_1,..,m_{j+1},m_j,..,m_n}=
-\varphi_{m_1,..,m_j,m_{j+1},..,m_n}+2(-1)^{m_j} \delta_{m_j+m_{j+1},0}
\varphi_{m_1,..,{\widehat {m_j}},{\widehat {m_{j+1}}},..,m_n}$.

\noindent{}The only non-trivial rule is the last one, which follows from the
identity
$$\sum_{0 \leq r \leq m} (-1)^r \alpha_r \alpha_{m-r} = \delta_{m,0}$$
Starting from some $\varphi_{m_1,..,m_n}$, to reexpress it in terms of
$\fd$'s one has to use the rule (iv) repeatedly to ``push" the negative indices
to the right, which results in either (ii) or (iii), and ends up with an
expression
of $f$ through (i).
Note that in this way $\phid$'s are expressed as linear
combinations of the $\fd$'s with relative integer coefficients.
We have for instance
\eqn\instphi{
\eqalign{
\varphi_{m_1,-m_2,m_3}&=2(-1)^{m_2} \delta_{m_2,m_3} \
f_{m_1} \ \ \ \ \ m_1,m_2,m_3 > 0 \cr
\varphi_{-m_1,m_2,m_3}&=2(-1)^{m_1} [\delta_{m_1,m_2}
\ f_{m_3} -
\delta_{m_1,m_3} \ f_{m_2}]\ \ m_1,m_2,m_3>0 \cr}}
which should be compared with \example.

We now turn to the actual computation of the $\fd$'s in terms of
$\Gthd$'s.
{}From the last equality in \plicit, we get a recursion relation
for $\fd$'s
\eqn\recurf{ f_{k_1,...,k_{n+1}}(\Gthd)=
\sum_{s=0}^{k_{n+1}} \  p_{k_{n+1}-s} \sum_{r_1,..,r_n \geq 0
\atop r_1+...+r_n = s}  \big( \prod_{i=1}^n \alpha_{r_i} \big)
f_{k_1+r_1,...,k_n+r_n}(\Gthd)}
Let us use the shorthand notation
\eqn\defthetnu{ \theta_{\{\nu\}} =\theta_{[1^{\nu_1}3^{\nu_3}...]}=
\prod_{odd \ j >0} {{ \theta_j^{\nu_j} }
\over \nu_j ! } }
for any set $\{ \nu \}=\nu_1,\nu_3,...$ of non negative integers (or
alternatively any permutation $[1^{\nu_1} 3^{\nu_3}...]$ of
$\sum_{odd \ j} j \nu_j$ elements with odd cycles only),
then the coefficients $A$ in the expansion
\eqn\fexpan{ f_{k_1,...,k_n}(\Gthd) = \sum_{\nu_j \geq 0,j \ odd \atop
\nu_1+3 \nu_3+...= k_1+...+k_n}
A_{k_1,...,k_n}^{\{ \nu \}} \theta_{\{ \nu \}} }
satisfy the recursion relation
\eqn\recurak{
\eqalign{
A_{k_1,...,k_{n+1}}^{\{ \nu \}}&=
\sum_{\mu_j \geq 0, j \ odd \atop
\mu_j \leq \nu_j; (\nu_1-\mu_1) +3(\nu_3-\mu_3)+...\leq k_{n+1}}
\prod_{j \ odd} { \nu_j \choose \mu_j}  \times \cr
&\times \sum_{r_1,...,r_n \geq 0 \atop
r_1+...+r_n=k_{n+1}+(\mu_1 - \nu_1)+3(\mu_3-\nu_3)+... }
\big( \prod_{i=1}^n \alpha_{r_i} \big)
A_{k_1+r_1,...,k_n+r_n}^{\{ \mu \}} \cr } }

Considering that only integer coefficients enter the recursion relation
and that $A_k^{\{ \nu \}}=1$ for $\sum_{j \ odd} j \nu_j =k$, we
deduce that all $A$'s are integers.
It is now easy to compute the first few $A$'s, we find (the
multi-index superscript
$\{\nu\}$ is always
related to the indices through $\sum_j j \nu_j=\sum k_i$)
\eqn\acomp{
\eqalign{
A_{k}^{\{\nu\}} &=1 \cr
A_{k,1}^{\{\nu\}}&= \nu_1-2 \cr
A_{k,2}^{\{\nu\}}&= {{(\nu_1-1)(\nu_1-4)} \over 2!} \cr
A_{k,3}^{\{\nu\}}&={{(\nu_1-1)(\nu_1-2)(\nu_1-6)}
\over 3!}+\nu_3 \cr
A_{k,4}^{\{\nu\}}&={{(\nu_1-1)(\nu_1-2)
(\nu_1-3)(\nu_1-8)}\over 4!}+(\nu_1-2)\nu_3 \cr
A_{k,5}^{\{\nu\}}&={{(\nu_1-1)(\nu_1-2)(\nu_1-3)
(\nu_1-4)(\nu_1-10)} \over 5!}
+{{(\nu_1-1)(\nu_1-4)}\over 2!}\nu_3 + \nu_5 \cr
A_{k,2,1}^{\{\nu\}}&={{\nu_1(\nu_1-4)(\nu_1-5)}
\over 3!}-2\nu_3 \cr
A_{k,3,1}^{\{\nu\}}&=2 {{ \nu_1(\nu_1-2)(\nu_1-5)
(\nu_1-7)} \over 4!} -(\nu_1-2)\nu_3 \cr
A_{k,4,1}^{\{\nu\}}&=3 {{ \nu_1(\nu_1-2)(\nu_1-3)
(\nu_1-6)(\nu_1-9)} \over 5!} -2 \nu_5 \cr
A_{k,3,2}^{\{\nu\}}&= 2{{\nu_1(\nu_1-1)(\nu_1-4)
(\nu_1-7)(\nu_1-8)} \over 5!}
-{{(\nu_1-1)(\nu_1-4)}\over 2}
\nu_3 + 2 \nu_5 \cr}}
Table I is obtained by using these expressions.

\vfill
\eject

%
%

\appendix {B} {Two illustrative cases}
\noindent
For any invariant polynomial $f(Y)$, {\it i.e.} depending only on the
eigenvalues of $Y$,
one may write following the steps of \CJB
\eqn\IIIa{\eqalign{\ll f(&Y+  i\GL)\gg
=\prod_i \Gl_i^{\oh}\prod_{i<j} (\Gl_i+\Gl_j) \int dY f(Y+i\GL) e^{{i\over 6}
\tr Y^3-\oh \tr Y^2\GL}\cr
&= \prod_i \Gl_i^{\oh}\prod_{i<j} (\Gl_i+\Gl_j) e^{{1\over 6}\tr \GL^3}
\int dY f(Y) e^{i\({1\over 6}\tr Y^3+\oh \tr Y\GL^2\)}\cr
&={1 \over N!} \int  \prod_i {dy_i \over \sqrt{2 \pi}}
f(y_j) e^{\inv {3}\sum \Gl_i^3+
i\sum {1\over 6}y_i^3+\oh y_i\Gl_i^2} \prod \Gl_i^{\oh}\prod_{i<j}
\({y_i-y_j\over \Gl_j-\Gl_i}\)\cr
&= {1 \over N!} \int  \prod_i {dy_i \over \sqrt{2 \pi}}
\(f\(-2i\dd{}{\Gl_j^2}\)
e^{i\sum {1\over 6}y_i^3+\oh y_i\Gl_i^2} \)
e^{\inv{3} \sum \Gl_i^3} \prod\Gl_i^{\oh}{\prod_{i<j}(y_i-y_j)\over
\GD(\Gl)}\cr
&= {1\over {N! \GD(\Gl)}}f\(i\Gl_j^{\oh} e^{{1\over 3} \Gl_j^3}
\Big(-2\dd{}{\Gl_j^2}\Big) \Gl_j^{-\oh} e^{-{1\over 3} \Gl_j^3}\)
\int\prod {dy_i \over \sqrt{2 \pi}} \Gl_i^{\oh}e^{{1\over 3} \Gl_i^3}
\prod_{i<j} (y_i-y_j) e^{i\sum {1\over 6}y_i^3+\oh y_i\Gl_i^2} \cr
&={1\over \GD(\Gl)} f(iD) \GD(\Gl) \Xi\ . \cr}}
where the double bracket denotes the weighted average
\avkon, $\GD(\Gl)$  the Vandermonde
determinant of the $\Gl$'s and
\eqn\IIIb{\eqalign{D_i &=\Gl_i+{2\over \Gl_i}-{1\over\Gl_i}\dd{}{\Gl_i}
\cr &= -e^{\inv{3}\Gl_i^3} \Gl_i^{\oh} {2\partial\over \partial \Gl_i^2}
\Gl_i^{-\oh} e^{-\inv{3} \Gl_i^3}\ .\cr }}
The second equality in \IIIa\ is obtained by a translation $Y \to Y-i\Lambda$
for an analytic continuation to pure imaginary $\GL$
(see section 3). The next one follows from an angular integration.

The above relation is in particular true for arbitrary powers of $\tr (Y+i\GL)$
\eqn\IIIc{\ll [\tr ({Y\over i}+\GL)]^p \gg =
{1\over \GD(\Gl)}\,[\tr\, D]^p \GD(\Gl)\, \Xi\ .}
One forms a generating function for the averages of
powers of $\tr Y$ in the form
\eqn\IIId{
\eqalign{
\ll e^{s \tr {Y\over i}} \gg
&= e^{-s\tr \GL} \ll e^{s \tr ({Y\over i}+\GL)} \gg \cr
&=e^{-s\tr \GL} {1\over \GD(\Gl)} e^{s\tr D} \GD(\Gl) \Xi \cr
&= e^{-s\tr \GL} \( {1\over \GD(\Gl)} e^{{1\over 3}\tr \GL^3} \det\GL^{\oh}\)
e^{-2s\sum\dd{}{\Gl_i^2}}
\(\det\GL^{-\oh}e^{-\inv{3}\tr \GL^3}\GD(\Gl) \Xi\)\ . \cr }}
The operator $e^{-2s\sum\dd{}{\Gl_i^2}} $ shifts all variables
$\Gl_i^2$ by $-2s$. Hence,
using the expressions displayed in \CJB, one finds after some algebra
\eqn\IIIe{\ll e^{s \tr {Y\over i}}\gg
=e^{\psi(s)}\exp s\sum_{n \geq 0} t_{n+1}\dd{}{t_n}\Xi}
where the variables $t_{.}$ are related to our $\Gthd$ defined
in \nort\ through
\eqn\relattet{
t_k ={ (2k+1)!! \over 2}  \theta_{2k+1}}
and $\psi$ is the function given by
\eqn\IIIf{\psi(s)= {1\over 3} \sum \Gl_i^3
-{1\over 3}\sum (\Gl_i^2-2s)^{\oh}+\sum_{i<j}\ln {\Gl_i+\Gl_j\over
\sqrt{\Gl_i^2-2s}+\sqrt{\Gl_j^2-2s}}
-{1\over 4}\ln {\Gl_i^2-2s\over \Gl_i^2}-s\sum \Gl_i\ . }
The differential operator $l_{-1}$ that appears in the exponential of the
r.h.s. of \IIIe\
is part of a Virasoro operator that annihilates the function
$\Xi$, namely
$L_{-1}=\sum_{ n \geq 0} t_{n+1}\dd{}{t_n}+\oh t_0^2-\dd{}{t_0}$.
This may be used to rewrite \IIIe\ as
\eqn\IIIfa{\ll e^{s \tr {Y\over i}}\gg =e^{\psi(s)} e^{sl_{-1}}e^{-sL_{-1}}
\Xi\ .}
Denoting $K(s)= e^{\psi(s)} e^{sl_{-1}}e^{-sL_{-1}}$, one finds that
$\dd{}{s}K(s)= (\dd{}{t_0} -\oh s^2) K(s)$ whence
\eqn\IIIg{ \ll e^{s \tr {Y\over i}}\gg
=e^{s\dd{}{t_0}-{s^3\over 6}} \Xi \ .}
or equivalently
\eqn\IIII{  e^{-2 s {\partial \over \partial \theta_1}} \Xi =
\ll e^{s \theta_1({Y \over 2i})-{s^3 \over 6}} \gg }
in terms of our $\Gthd$'s.
Therefore in this particular case, we have a very explicit
expression of correlation functions in terms of derivatives of $\Xi$,
vindicating the general proposition (W).

The averages of powers of $\tr Y^3$ may also be treated
simply. By a rescaling of the integration
variable $Y$, it is easy to derive
\eqn\IIIh{
\ll e^{y \tr\( {Y\over 2i}\)^3}\gg
= e^{{2\over 3}\ln (1+(3y/4)) \sum (2n+1)t_n\dd{}{t_n}}\Xi\ . }
Here too, the differential operator $l_0=\sum (2n+1) t_n\dd{}{t_n}$
in the exponential is a part
of the Virasoro generator $L_0=l_0+{1\over 16}-{3\over 2}\dd{}{t_1}$
that annihilates $\Xi$.
As above, we define $K'(z)=e^{2zl_0} e^{-2zL_0}$ and compute that
$$ \dd{}{z} K'(z)= \(3e^{-3z}\dd{}{t_1}- {1\over 8}\)K'(z)$$
Thus $K'(z)=\exp{-[(e^{-3z}-1)\dd{}{t_1}+{z\over 8}]}$.
This leads to
\eqn\IIIi{\ll e^{y \tr\( {Y\over 2i}\)^3}\gg
=(1+(3y/4))^{-{1\over 12}}e^{-[(1+(3y/4))^{-2}-1]\dd{}{t_1}}\Xi}
showing again that any
$\ll [\tr\( {Y\over 2i}\)^3]^p \gg $ is given by a polynomial in
the derivative $\partial \over {\partial t_1}$
acting on $\Xi$. The two special cases \IIIg\ and \IIIi\
may be combined into
\eqn\lafin{\ll e^{s\tr {Y\over i}
+y \tr\( {Y\over 2i}\)^3}\gg =
\exp\(\tilde s \dd{}{\tilde t_0} -{\tilde s^3\over 6}\)\, \Xi(\tilde \GL^{-1})}
where $\tilde \GL=(y/8)^{-{2\over 3}} \GL$, hence $\tilde t_n=(y/8)^{{2\over 3}
(2n+1) }t_n$ and $\tilde s= (y/8)^{-\inv{3} }s$.

\vfill
\eject

\def\gl{\lambda}
\def\gL{\Lambda}
\def\d{ d}
\def\tr{{\rm tr}}
\def\c{{\rm ch}}

%
\def\frac#1#2{{\scriptstyle{#1 \over #2}}}		
\def\inv#1{{\scriptstyle{1 \over #1}}}

\def\d{\partial}

\def\dd#1#2{{\partial #1 \over \partial #2}}

%
%

%

%


\def\IR{\relax{\rm I\kern-.18em R}}
\font\cmss=cmss10 \font\cmsss=cmss10 at 7pt
\def\IZ{\relax\ifmmode\mathchoice
{\hbox{\cmss Z\kern-.4em Z}}{\hbox{\cmss Z\kern-.4em Z}}
{\lower.9pt\hbox{\cmsss Z\kern-.4em Z}}
{\lower1.2pt\hbox{\cmsss Z\kern-.4em Z}}\else{\cmss Z\kern-.4em Z}\fi}
\def\inbar{\,\vrule height1.5ex width.4pt depth0pt}
\def\IB{\relax{\rm I\kern-.18em B}}
\def\IC{\relax\hbox{$\inbar\kern-.3em{\rm C}$}}
\def\ID{\relax{\rm I\kern-.18em D}}
\def\IE{\relax{\rm I\kern-.18em E}}
\def\IF{\relax{\rm I\kern-.18em F}}
\def\IG{\relax\hbox{$\inbar\kern-.3em{\rm G}$}}
\def\IH{\relax{\rm I\kern-.18em H}}
\def\II{\relax{\rm I\kern-.18em I}}
\def\IK{\relax{\rm I\kern-.18em K}}
\def\IL{\relax{\rm I\kern-.18em L}}
\def\IM{\relax{\rm I\kern-.18em M}}
\def\IN{\relax{\rm I\kern-.18em N}}
\def\IO{\relax\hbox{$\inbar\kern-.3em{\rm O}$}}
\def\IP{\relax{\rm I\kern-.18em P}}
\def\IQ{\relax\hbox{$\inbar\kern-.3em{\rm Q}$}}
\def\IGa{\relax\hbox{${\rm I}\kern-.18em\Gamma$}}
\def\IPi{\relax\hbox{${\rm I}\kern-.18em\Pi$}}
\def\ITh{\relax\hbox{$\inbar\kern-.3em\Theta$}}
\def\IOm{\relax\hbox{$\inbar\kern-3.00pt\Omega$}}


\def\d{{\rm d}}

\def\oh{{1\over 2}}


\def\GD{\Delta}
\def\Gth{\theta}
\def\Gl{\lambda}\def\GL{\Lambda}

\def\GS{\Sigma}


\def\d{{\rm d}}

\def\Gthd{\Gth_{\textstyle .} }

\def\K{Kontsevich}
\def\tvi{\vrule height 12pt depth 6pt width 0pt}\def\tv{\tvi\vrule}
\centerline{ \bf Table I}
$$\vbox{\offinterlineskip
\halign{\tv\quad # & # \hfill &\quad \tv#\cr    
\noalign{\hrule}
\tvi $f_1$ &$= \theta_{[1^1]}$& \cr
\noalign{\hrule}
\tvi $f_2$ &$= \theta_{[1^2]}$ &\cr
\noalign{\hrule}
\tvi $f_3$ &$= \theta_{[1^3]}+\theta_{[3^1]}$ &\cr
\tvi $f_{2,1}$ &$= \theta_{[1^3]} -2 \theta_{[3^1]} $&\cr
\noalign{\hrule}
\tvi $f_4$&$= \theta_{[1^4]} + \theta_{[1^1 3^1]}$ &\cr
\tvi $f_{3,1}$ &$= 2\theta_{[1^4]} -\theta_{[1^1 3^1]}$ &\cr
\noalign{\hrule}
\tvi $f_5$&$= \theta_{[1^5]} + \theta_{[1^2 3^1]} +\theta_{[5^1]}$ &\cr
\tvi $f_{4,1}$&$= 3\theta_{[1^5]} - 2 \theta_{[5^1]}$ &\cr
\tvi $f_{3,2}$&$= 2 \theta_{[1^5]} - \theta_{[1^2 3^1]} + 2 \theta_{[5^1]}$&\cr
\noalign{\hrule} 
\tvi $f_6$ &$= \theta_{[1^6]}+ \theta_{[1^3 3^1]}+\theta_{[1^1 5^1]}
+\theta_{[3^2]}$ &\cr
\tvi $f_{5,1}$&$= 4\theta_{[1^6]}+ \theta_{[1^3 3^1]}-\theta_{[1^1 5^1]}
-2 \theta_{[3^2]}$ &\cr
\tvi $f_{4,2}$&$= 5\theta_{[1^6]} - \theta_{[1^3 3^1]}+2 \theta_{[3^2]}$ &\cr
\tvi $f_{3,2,1}$&$= 2 \theta_{[1^6]}-\theta_{[1^3 3^1]}
+2 \theta_{[1^1 5^1]} -4 \theta_{[3^2]} $&\cr
\noalign{\hrule}
\tvi $f_7$ &$= \theta_{[1^7]}+\theta_{[1^4 3^1]}+\theta_{[1^2 5^1]}
+\theta_{[1^1 3^2]}+ \theta_{[7^1]}$ &\cr
\tvi $f_{6,1}$&$= 5 \theta_{[1^7]}+2 \theta_{[1^4 3^1]}
-\theta_{[1^1 3^2]} -2 \theta_{[7^1]}$ &\cr
\tvi$ f_{5,2}$&$= 9 \theta_{[1^7]} - \theta_{[1^2 5^1]}
+2 \theta_{[7^1]}$ &\cr
\tvi $f_{4,3}$&$= 5 \theta_{[1^7]} - \theta_{[1^4 3^1]}
+2 \theta_{[1^1 3^2]} -2\theta_{[7^1]}$&\cr
\tvi $f_{4,2,1}$&$= 7 \theta_{[1^7]}- 2\theta_{[1^4 3^1]}
+2 \theta_{[1^2 5^1]} -2 \theta_{[1^1 3^2]}$ &\cr
\noalign{\hrule}
\tvi $f_8$ &$= \theta_{[1^8]} + \theta_{[1^5 3^1]}
+\theta_{[1^3 5^1]} +\theta_{[1^2 3^2]}
+\theta_{[1^1 7^1]}+\theta_{[3^1 5^1]}$ &\cr
\tvi $f_{7,1}$&$= 6 \theta_{[1^8]}+3 \theta_{[1^5 3^1]}
+\theta_{[1^3 5^1]}- \theta_{[1^1 7^1]}
-2 \theta_{[3^1 5^1]}$ &\cr
\tvi $f_{6,2}$&$= 14 \theta_{[1^8]} + 2 \theta_{[1^5 3^1]}
-\theta_{[1^3 5^1]} -\theta_{[1^2 3^2]}
+2 \theta_{[3^1 5^1]}$ &\cr
\tvi $f_{5,3}$&$=14 \theta_{[1^8]} -\theta_{[1^5 3^1]}
-\theta_{[1^3 5^1]} + 2 \theta_{[1^2 3^2]}
-\theta_{[3^1 5^1]}$ &\cr
\tvi $f_{5,2,1}$&$= 16 \theta_{[1^8]} -2 \theta_{[1^5 3^1]}
+\theta_{[1^3 5^1]} -2 \theta_{[1^2 3^2]}
+2 \theta_{[1^1 7^1]} -2 \theta_{[3^1 5^1]}$&\cr
\tvi $f_{4,3,1}$&$=12 \theta_{[1^8]}-3\theta_{[1^5 3^1]}
+2 \theta_{[1^3 5^1]} -2\theta_{[1^1 7^1]}
+2 \theta_{[3^1 5^1]} $ &\cr
\noalign{\hrule} }} $$

The $f$ polynomials up to degree 8. The notation
$\theta_{[1^{\nu_1} 3^{\nu_3}...(2k+1)^{\nu_{2k+1}}..]}$
is a short hand for ${\theta_1^{\nu_1} \over \nu_1 !}
{\theta_3^{\nu_3} \over \nu_3 !} .. {\theta_{2k+1}^{\nu_{2k+1}} \over
\nu_{2k+1} !} ...$.
\vfill
\eject

\centerline{ \bf Table II}

$$\vbox{\offinterlineskip
\halign{\tv\quad # & # \hfill &\quad \tv#\cr    
\noalign{\hrule}
$\theta_{[1^1]} $&$= f_1 $&\cr
\noalign{\hrule}
$\theta_{[1^2]}$&$= f_2 $&\cr
\noalign{\hrule}
$\theta_{[1^3]}$&$=\inv{3}[2f_3+f_{2,1}] $&\cr
$\theta_{[3^1]}$&$=\inv{3}[f_3 -f_{2,1}] $&\cr
\noalign{\hrule}
$\theta_{[1^4]}$&$=\inv{3}[f_4+f_{3,1}] $&\cr
$\theta_{[1^1 3^1]}$&$=\inv{3}[2f_4 -f_{3,1}]  $&\cr
\noalign{\hrule}
$\theta_{[1^5]}$&$=\inv{15}[2f_5+3 f_{4,1}+2f_{3,2}] $&\cr
$\theta_{[1^2 3^1]}$&$=\inv{3}[2f_5-f_{3,2}]$&\cr
$\theta_{[5^1]}$&$=\inv{5}[f_5-f_{4,1}+f_{3,2}]$&\cr
\noalign{\hrule}
$\theta_{[1^6]}$&$=\inv{45}[2f_6+4f_{5,1}+5f_{4,2}
+f_{3,2,1}]$&\cr
$\theta_{[1^3 3^1]}$&$=\inv{9}[4f_6+2f_{5,1}-2f_{4,2}
-f_{3,2,1}]$&\cr
$\theta_{[1^1 5^1]}$&$=\inv{5}[2f_6-f_{5,1}+f_{3,2,1}]$&\cr
$\theta_{[3^2]}$&$=\inv{9}[f_6-f_{5,1}+f_{4,2}-f_{3,2,1}]$&\cr
\noalign{\hrule}
$\theta_{[1^7]}$&$=\inv{315}[4f_7+10f_{6,1}
+18f_{5,2}+10f_{4,3}+7f_{4,2,1}]$&\cr
$\theta_{[1^4 3^1]}$&$=\inv{9}[2f_7+2f_{6,1}-f_{4,3}
-f_{4,2,1}]$&\cr
$\theta_{[1^2 5^1]}$&$=\inv{5}[2f_7-f_{5,2}+f_{4,2,1}]$&\cr
$\theta_{[1^1 3^2]}$&$=\inv{9}[2f_7-f_{6,1}+2f_{4,3}
-f_{4,2,1}]$&\cr
$\theta_{[7^1]}$&$=\inv{7}[f_7+f_{5,2}
-f_{6,1}-f_{4,3}]$&\cr
\noalign{\hrule}
$\theta_{[1^8]}$&$=\inv{315}[f_8+3f_{7,1}+7f_{6,2}
+7f_{5,3}+4f_{5,2,1}+3f_{4,3,1}]$&\cr
$\theta_{[1^5 3^1]}$&$=\inv{45}[4f_8+6f_{7,1}+4f_{6,2}
-2f_{5,3}-2f_{5,2,1}-3f_{4,3,1}]$&\cr
$\theta_{[1^3 5^1]}$&$=\inv{15}[4f_8+2f_{7,1}-2f_{6,2}
-2f_{5,3}+f_{5,2,1}+2f_{4,3,1}]$&\cr
$\theta_{[1^2 3^2]}$&$=\inv{9}[2f_8-f_{6,2}+2f_{5,3}
-f_{5,2,1}]$&\cr
$\theta_{[1^1 7^1]}$&$=\inv{7}[2f_8 -f_{7,1}+f_{5,2,1}
-f_{4,3,1}]$&\cr
$\theta_{[3^1 5^1]}$&$=\inv{15}[2f_8-2f_{7,1}+2f_{6,2}
-f_{5,3}-f_{5,2,1}+f_{4,3,1}]$&\cr
\noalign{\hrule} }} $$

Expression of $\theta_{\textstyle .}$ monomials in terms of
$f_{\textstyle .}$'s up to degree 8.

\vfill
\eject

\def\p{{\partial_{\scriptstyle .}  }}

\def\pt{{\partial_{\theta_{\scriptstyle .}}  }}

\def\tvi{\vrule height 12pt depth 6pt width 0pt}\def\tv{\tvi\vrule}
\centerline{ \bf Table III}
$$\vbox{\offinterlineskip
\halign{\tv\quad # & # \hfill &\quad \tv#\cr    
\noalign{\hrule}
\tvi $f_1(\p ) \Xi $ &$=\ \ll f_1 \gg $& \cr
\noalign{\hrule}
\tvi $f_2(\p )\Xi $ &$=\ \ll f_2 \gg $ &\cr
\noalign{\hrule}
\tvi $f_3(\p )\Xi $ &$=\ \ll f_3 -{5 \over 24} \gg $ &\cr
\tvi $f_{2,1}(\p )\Xi $ &$= \ll f_{2,1} -{1 \over 12} \gg $&\cr
\noalign{\hrule}
\tvi $f_4(\p )\Xi $&$=\ \ll f_4 -{17 \over 24} f_1 \gg $ &\cr
\tvi $f_{3,1}(\p )\Xi $ &$=\ \ll f_{3,1} +{5 \over 24} f_1 \gg $ &\cr
\noalign{\hrule}
\tvi $f_5(\p )\Xi $&$=\ \ll f_5 -{35 \over 24} f_2 \gg $ &\cr
\tvi $f_{4,1}(\p )\Xi $&$=\ \ll f_{4,1} \gg $ &\cr
\tvi $f_{3,2}(\p )\Xi $&$=\ \ll f_{3,2} +{5 \over 24} f_2 \gg $&\cr
\noalign{\hrule} 
\tvi $f_6(\p )\Xi $ &$=\ \ll f_6 -{59 \over 24} f_3 + {385 \over 1152} \gg $
&\cr
\tvi $f_{5,1}(\p )\Xi $&$=\ \ll f_{5,1} -{35 \over 24} f_{2,1} +{35 \over 576}
\gg $ &\cr
\tvi $f_{4,2}(\p )\Xi $&$=\ \ll f_{4,2} +{17 \over 24} f_{2,1} -{35 \over 576}
\gg $ &\cr
\tvi $f_{3,2,1}(\p )\Xi $&$=\ \ll f_{3,2,1} - {5 \over 24} f_{2,1} -{1 \over
12} f_3
+ {5 \over 288} \gg  $&\cr
\noalign{\hrule}
\tvi $f_7(\p )\Xi $ &$=\ \ll f_7 -{89 \over 24} f_4 + {1801 \over 1152}f_1 \gg
$ &\cr
\tvi $f_{6,1}(\p )\Xi $&$=\ \ll f_{6,1} -{ 59 \over 24} f_{3,1}
-{385 \over 1152} f_1 \gg $ &\cr
\tvi$ f_{5,2}(\p )\Xi $&$= \ \ll  f_{5,2} \gg $ &\cr
\tvi $f_{4,3}(\p )\Xi $&$=\ \ll f_{4,3} + { 17 \over 24} f_{3,1}
-{ 5 \over 24} f_4 + { 85 \over 576} f_1 \gg $&\cr
\tvi $f_{4,2,1}(\p )\Xi $&$=\ \ll f_{4,2,1} -{ 1 \over 12} f_4
-{ 1 \over 576} f_1 \gg $ &\cr
\noalign{\hrule}
\tvi $f_8(\p )\Xi $ &$=\ \ll f_8 -{ 125 \over 24} f_5
+{5005 \over 1152} f_2 \gg $ &\cr
\tvi $f_{7,1}(\p )\Xi $&$=\ \ll f_{7,1} -{89 \over 24} f_{4,1} \gg $ &\cr
\tvi $f_{6,2}(\p )\Xi $&$=\ \ll f_{6,2} -{59 \over 24} f_{3,2}
-{385 \over 1152} f_2 \gg $ &\cr
\tvi $f_{5,3}(\p )\Xi $&$=\ \ll f_{5,3} + {35 \over 24} f_{3,2}
-{ 5 \over 24} f_5 + {175 \over 576} f_2 \gg $ &\cr
\tvi $f_{5,2,1}(\p )\Xi $&$=\ \ll f_{5,2,1} -{1 \over 12} f_5 +
{35 \over 576} f_2 \gg $&\cr
\tvi $f_{4,3,1}(\p )\Xi $&$= \ \ll f_{4,3,1} +{5 \over 24} f_{4,1} \gg $ &\cr
\noalign{\hrule} }} $$

The derivatives of the Kontsevich partition function with respect to the
$\theta_{\textstyle .}$'s expressed as averages over polynomials in odd traces.
The notation $\partial_{\textstyle .}$ stands for $\{-{2 \over 2k+1} {\partial
\over { \partial \theta_{2k+1}}} \}$, $\theta_{\textstyle_ .}\equiv
\theta(\gL^{-1})$,
while on the r.h.s. the matrix argument of the $f_{\scriptstyle .}$'s is
$Y/2i$.

\vfill
\eject

\centerline{ \bf Table IV}

$$\vbox{\offinterlineskip
\halign{\tv\quad # & # \hfill &\quad \tv#\cr    
\noalign{\hrule}
$\theta_{[1^1]}(\p ) \Xi $&$= \ll \theta_{[1^1]} \gg $&\cr
\noalign{\hrule}
$\theta_{[1^2]}(\pt ) \Xi $&$= \ll \theta_{[1^2]} \gg $&\cr
\noalign{\hrule}
$\theta_{[1^3]}(\pt ) \Xi $&$=\ll \theta_{[1^3]} -{1 \over 6} \gg $&\cr
$\theta_{[3^1]}(\pt ) \Xi $&$=\ll \theta_{[3^1]}-\inv{24} \gg $&\cr
\noalign{\hrule}
$\theta_{[1^4]}(\pt ) \Xi $&$=\ll \theta_{[1^4]}
- \inv{6} \theta_{[1^1]}\gg $&\cr
$\theta_{[1^1 3^1]}(\pt ) \Xi $&$=\ll \theta_{[1^1 3^1]}
-{13 \over 24} \theta_{[1^1]}\gg $&\cr
\noalign{\hrule}
$\theta_{[1^5]}(\pt ) \Xi $&$=\ll \theta_{[1^5]}
- \inv{6} \theta_{[1^2]}\gg $&\cr
$\theta_{[1^2 3^1]}(\pt ) \Xi $&$=\ll \theta_{[1^2 3^1]}
-{25 \over 24}  \theta_{[1^2]}\gg $&\cr
$\theta_{[5^1]}(\pt ) \Xi $&$=\ll \theta_{[5^1]}
-{1 \over 4} \theta_{[1^2]} \gg $&\cr
\noalign{\hrule}
$\theta_{[1^6]}(\pt ) \Xi $&$=\ll \theta_{[1^6]} -\inv{6}
\theta_{[1^3]} + \inv{72} \gg $&\cr
$\theta_{[1^3 3^1]}(\pt ) \Xi $&$=\ll \theta_{[1^3 3^1]}
-{37 \over 24} \theta_{[1^3]} -\inv{6} \theta_{[3^1]}
+{25 \over 144} \gg $&\cr
$\theta_{[1^1 5^1]}(\pt ) \Xi $&$=\ll \theta_{[1^1 5^1]}
-{3 \over 4} \theta_{[1^3]} -{3 \over 2}
\theta_{[3^1]} +\inv{8} \gg $&\cr
$\theta_{[3^2]}(\pt ) \Xi $&$=\ll \theta_{[3^2]} -{19 \over 24}
\theta_{[3^1]} +{25 \over 1152} \gg $&\cr
\noalign{\hrule}
$\theta_{[1^7]}(\pt ) \Xi $&$=\ll \theta_{[1^7]} -\inv{6}
\theta_{[1^4]} +\inv{72} \theta_{[1^1]}\gg $&\cr
$\theta_{[1^4 3^1]}(\pt ) \Xi $&$=\ll \theta_{[1^4 3^1]}
-{49 \over 24}\theta_{[1^4]} -\inv{6} \theta_{[1^1 3^1]}
+{37 \over 144}  \theta_{[1^1]} \gg $&\cr
$\theta_{[1^2 5^1]}(\pt ) \Xi $&$=\ll \theta_{[1^2 5^1]}
-{3 \over 2} \theta_{[1^4]} -{3 \over 2} \theta_{[1^1 3^1]}
+{5 \over 8}  \theta_{[1^1]}\gg $&\cr
$\theta_{[1^1 3^2]}(\pt ) \Xi $&$=\ll \theta_{[1^1 3^2]} -{31 \over 24}
\theta_{[1^1 3^1]} +{481 \over 1152}  \theta_{[1^1]}
\gg $&\cr
$\theta_{[7^1]}(\pt ) \Xi $&$=\ll \theta_{[7^1]} -{3 \over 4}
\theta_{[1^1 3^1]} + \inv{5}  \theta_{[1^1]} \gg $&\cr
\noalign{\hrule}
$\theta_{[1^8]}(\pt ) \Xi $&$=\ll \theta_{[1^8]} -\inv{6} \theta_{[1^5]}
 +\inv{72} \theta_{[1^2]} \gg $&\cr
$\theta_{[1^5 3^1]}(\pt ) \Xi $&$=\ll \theta_{[1^5 3^1]} -{61 \over 24}
\theta_{[1^5]}-\inv{6} \theta_{[1^2 3^1]}+{49 \over 144}  \theta_{[1^2]}
\gg $&\cr
$\theta_{[1^3 5^1]}(\pt ) \Xi $&$=\ll \theta_{[1^3 5^1]}
-{5 \over 2}
\theta_{[1^5]}
-{3 \over 2}  \theta_{[1^2 3^1]}-\inv{6} \theta_{[5^1]}+
{7 \over 6} \theta_{[1^2]}  \gg $&\cr
$\theta_{[1^2 3^2]}(\pt ) \Xi $&$=\ll \theta_{[1^2 3^2]}
-{43 \over 12} \theta_{[1^2 3^1]}
+{1225 \over 1152}  \theta_{[1^2]} \gg $&\cr
$\theta_{[1^1 7^1]}(\pt ) \Xi $&$=\ll \theta_{[1^1 7^1]}
-{3 \over 2} \theta_{[1^2 3^1]}-{5 \over 2} \theta_{[5^1]}
+{5 \over 4} \theta_{[1^2]} \gg $&\cr
$\theta_{[3^1 5^1]}(\pt ) \Xi $&$=\ll \theta_{[3^1 5^1]}
-{1 \over 4}  \theta_{[1^2 3^1]}-{61 \over 24}  \theta_{[5^1]}
+{49 \over 96}  \theta_{[1^2]} \gg $&\cr
\noalign{\hrule} }} $$
%

Monomials in the derivatives acting on the Kontsevich integral
expressed as averages of polynomials. The notation
$\theta_{[...(2k+1)^{\nu_{2k+1}}...]}(\pt )$ stands for
$...{1\over \nu_{2k+1} !} \partial_{2k+1}^{\nu_{2k+1} }...
\equiv ... {1 \over \nu_{2k+1} !}
{\big(-(2/2k+1)\partial_{\theta_{2k+1}}\big)^{\nu_{2k+1}}}...$.

\listrefs

\end